\documentclass{article}
\usepackage[utf8]{inputenc}

\usepackage[utf8]{inputenc}
\usepackage[left=3cm,right=3cm,top=2cm,bottom=2cm]{geometry}

\usepackage[english]{babel}
\usepackage[utf8]{inputenc}
\usepackage[T1]{fontenc}
\usepackage{amsmath,amsthm,amssymb}
\usepackage{mathrsfs}
\usepackage{caption}
\usepackage{mathtools} 
\usepackage{faktor} 
\usepackage{enumitem} 
\usepackage{etoolbox}
\usepackage{bbm}
\usepackage{hyperref}
\usepackage{titlesec}

\usepackage{graphicx}

\DeclareGraphicsExtensions{.png}

\theoremstyle{plain}

\newtheorem*{thm*}{Teorema}
\newtheorem*{lem*}{Lemma}
\newtheorem*{prop*}{Proposizione}

\theoremstyle{definition}

\newcommand{\rar}{\Longrightarrow}

\newcommand{\field}[1]{\mathbb{#1}}

\providecommand{\R}{\field{R}}

\providecommand{\de}{\mathrm{d}}

\usepackage{titling}
\newcommand{\subtitle}[1]{%
  \posttitle{%
    \par\end{center}
    \begin{center}\large#1\end{center}}%
}

\title{\textbf{Liquidity Providers Greeks and Impermanent Gain}}
\author{Niccolò Bardoscia, Alessandro Nodari \thanks{The authors are greteful to 0xSami\_, Andrea Bugin, Andrea Prampolini, Barry Fried, Carlo Sala, Fabio Bellini, Giulio Anselmi and Miguel Ottina for providing valuable feedback on drafts of the article.}}
\date{23 February 2023}
\begin{document}
\maketitle

\vspace{25pt}

\section*{Abstract}

In traditional finance, the Black \& Scholes model has guided almost 50 years of derivatives pricing, defining a standard to model any volatility-based product. With the rise of Decentralized Finance (DeFi) and constant product Automated Market Makers (AMMs), Liquidity Providers (LPs) are playing an increasingly important role in markets functioning, but, as the recent bear market highlighted, they are exposed to important risks such as Impermanent Loss (IL).

In this paper, we tailor the formulas introduced by Black \& Scholes to DeFi, proposing a method to calculate the greeks of an LP.
We also introduce Impermanent Gain, a product that LPs can use to hedge their position and traders can use to bet on a rise in volatility and benefit from large market moves.

\vspace{15pt}

\textit{Keywords: liquidity providers, impermanent loss, constant product automated market makers, impermanent gain}

\newpage

\section{Introduction}

Before diving deeper on the subjects of this article we will present, in this first section, a brief review of the ecosystem in which we will operate and of its principal actors.

\subsection{Decentralized Exchanges}

A Decentralized Exchange (DEX) is a peer-to-peer marketplace where transactions occur directly between crypto traders. DEXs fulfill one of crypto's core possibilities: fostering financial transactions that aren't officiated by banks, brokers, payment processors, or any other kind of intermediary. The most popular DEXs, like Uniswap and Sushiswap, utilize the Ethereum blockchain and are part of the growing suite of DeFi tools which make a huge range of financial services available directly from a compatible crypto wallet. Unlike Centralized Exchanges (CEXs) like Binance and Coinbase, DEXs don’t allow for exchanges between fiat and crypto.
All the transactions in a CEX are handled by the exchange itself via an order book that establishes the price for a particular cryptocurrency based on current buy and sell orders. DEXs, on the other hand, are simply a set of smart contracts: they establish the prices of various cryptocurrencies against each other algorithmically and use Liquidity Pools, in which investors lock funds in exchange for a percentage of the trading fees, in order to facilitate trades. While transactions on a CEX are recorded on that exchange’s internal database, DEX transactions are settled directly on the blockchain. DEXs are usually built on open-source code and developers can adapt existing code to create new competing projects.

\subsection{Liquidity Provider}

A Liquidity Provider in crypto, from here on LP, is an investor (individual or institution) who, as the name suggests, funds a liquidity pool with crypto assets it owns in order to facilitate trading on the platform and earn passive income on its deposit. The more assets in a pool, the more liquidity the pool has, and the easier trading becomes on a DEX for other market participants hence the crucial role played by LPs. How much the DEX pays the LP is based on the percentage of the crypto liquidity pool it puts, the volume, and the transaction fee offered by the exchange to LPs.

\subsubsection{Liquidity Provider Tokens}

Liquidity Provider tokens (LP tokens) are crypto tokens given to users who deposit their crypto into a Liquidity Pool. LP tokens represent the Liquidity Provider’s share of the pool and can be redeemed at any time for the underlying assets.
Some platforms require LP tokens to be locked for a period of time in order to access additional rewards (Liquidity Mining).

\subsection{Constant product AMM}

Here we will present in brief what is a constant product AMM for a more in depht analysis of AMM see for example [1], [2], [3] and [4]. A constant product AMM is a type of AMM in which the reserves of tokens are regulated by a product function, given two tokens $x$ and $y$ we have that:
$$x\cdot y=k$$
Where $x$ and $y$ represent, with an abuse of notation, the quantities of token $x$ and $y$ respectively and $k$ is the constant. For example in Uniswap, see [5] and [6], they call that constant $L^2$ so that:
$$\sqrt{x\cdot y}=L$$

We will indicate with $S_t$ the pool price of token $x$ in terms of token $y$ at time $t$. In this type of AMM we can derive the price as the ratio of the number of token $y$ and of token $x$ at time $t$:

$$S_t=\frac{y_t}{x_t}$$

Another interesting feature of this type of pools is that knowing in an instant $t_0$ the following data:

\begin{itemize}
    \item the constant of the pool $L$;
    \item the number of token $x$ at time $t_0$ given by $x_0$;
    \item the number of token $y$ at time $t_0$ given by $y_0$;
\end{itemize}

and supposing that there is no injection of new capital in the pool till time $T$, we can calculate the number of tokens in every time $t\in[t_0,T]$ as:

$$x_t=x_0\sqrt{\frac{S_0}{S_t}}\hspace{40pt}y_t=y_0\sqrt{\frac{S_t}{S_0}}\hspace{40pt}(1.3.1)$$

Having established these properties we can start analyzing how the position of a LP evolves in the pool. At time $t_0$, that we will assume without losing of generality that is equal to 0, he deposits a certain quantity of token $x$ and $y$ given by:
$$x_0=\frac{L}{\sqrt{S_0}}\hspace{50pt}y_0=L\sqrt{S_0}$$

Doing so we have that the constant product is respected. So we can calculate the initial capital invested, in terms of token $y$, as:

$$V_0 = x_0S_0+y_0=2L\sqrt{S_0}$$

When the price moves the quantity of each token changes according to (1.3.1) so that we have:

$$x_t=\frac{L}{\sqrt{S_t}}\hspace{50pt}y_t=L\sqrt{S_t}$$

And so it also changes the value of the position:
$$V_{LP}(t)=x_tS_t+y_t=2L\sqrt{S_t}=V_0\sqrt{\frac{S_t}{S_0}}=V_0\sqrt{\alpha_t}$$

Where we have defined $\alpha_t$ as the ratio of the change in price. So the P\&L of the LP position at time $t$ is:
$$LP(t)=V_t-V_0=V_0(\sqrt{\alpha_t}-1)$$

What we have ignored till now are the fees that Traders have to pay when they do their swaps. Part of these fees go to the protocol while the remainder is divided between the LPs proportional to the quantity of liquidity they have provided. The fees can be expressed as:
$$\Phi(t) = V_0\cdot \phi\cdot t$$

Where $\phi$ is the expected APY on the specific Liquidity Pool so that the actual payoff should be:
$$V_{LP}(t)=V_0\sqrt{\alpha_t}+\Phi(t)=V_0(\sqrt{\alpha_t}+\phi t)\hspace{20pt}(1.3.2)$$

While the final P\&L is:
$$LP(t)=V_0(\sqrt{\alpha_t}+\phi t-1)$$

\subsection{HODLer}

Before moving on we introduce one last character: the HODLer. In DeFi jargon an HODLer indicates a user that holds its tokens without doing anything. That is to say that given an initial quantity of tokens:

$$x_0\hspace{25pt}and\hspace{25pt}y_0$$

at every time $t$ the HODLer will have:

$$x_t=x_0\hspace{25pt}and\hspace{25pt}y_t=y_0$$

Thus the position value of an HODLer is determined just by the change in price of the tokens. The position value of an HODLer expressed in terms of token $y$ is given by:

$$V_H(t)=x_0S_t+y_0$$

\newpage

\section{Impermanent Loss}

Impermanent Loss (IL) is a known problem affecting the LPs defined as the difference between the return of a LP and that of an equal-weight (with respect to the starting time) HODLer. For a more in depth analysis of it on UniSwap, the most used constant product AMM, see [7] and [8]. We recall the starting quantities of tokens for a LP and hence of an equal-weight HODLer:

$$x_0=\frac{L}{\sqrt{S_0}}\hspace{50pt}y_0=L\sqrt{S_0}$$

and the values of a LP position and that of an HODLer at time $t$:

$$V_{LP}(t)=x_tS_t+y_t=\frac{LS_t}{\sqrt{S_t}}+L\sqrt{S_t}=V_0\sqrt{\alpha_t}$$

$$V_H(t)=x_0S_t+y_0=\frac{LS_t}{\sqrt{S_0}}+L\sqrt{S_0}=V_0\bigg(\frac{\alpha_t+1}{2}\bigg)$$

Now we can compute the IL:
$$IL(t)=\frac{V_{LP}(t)-V_0}{V_0}-\frac{V_H(t)-V_0}{V_0}=\frac{V_{LP}(t)-V_H(t)}{V_0}=\sqrt{\alpha_t}-\frac{\alpha_t}{2}-\frac{1}{2}\hspace{25pt}(2.1)$$

Formula (2) is the same found in [8]. We can also express that in terms of the token return defining:
$$r_t\coloneqq\frac{S_t}{S_0}-1=\alpha_t-1$$

Doing so we obtain the following equivalent form:
$$IL(r)=\sqrt{r+1}-\frac{r}{2}-1\hspace{25pt}(2.2)$$

\begingroup
\begin{center}
\includegraphics[scale=.38]{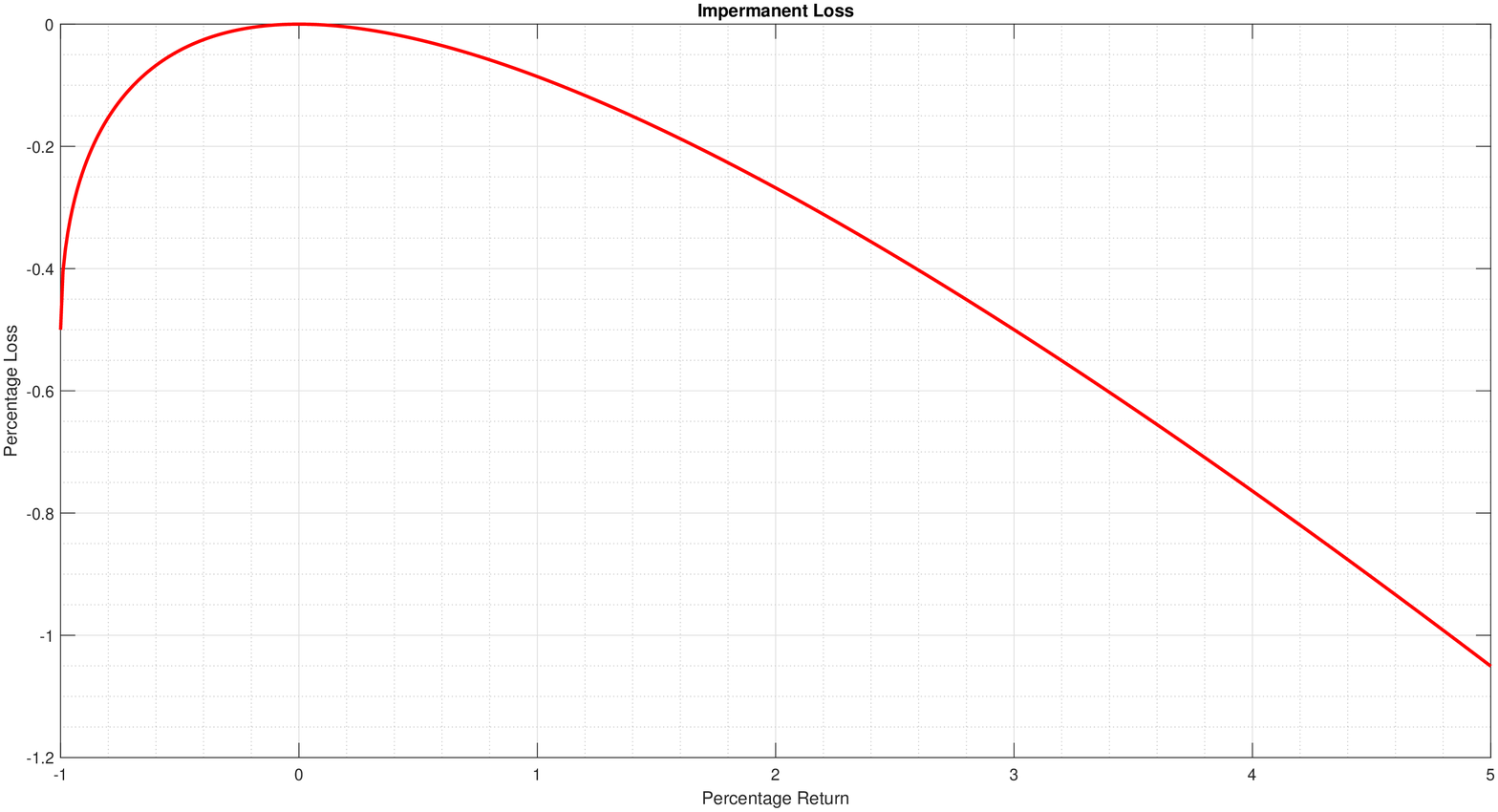}
\end{center}
\endgroup

We plotted the IL given by formula (2.2). We note that $IL(r)\leq0$ and equal to zero only if the price of token $x$ is exactly the same as its starting price $S_0$ ($r=0$). This tells us that being a LP is always worse than being an HODLer, unless the fees are enough to offset this difference.

\subsection{LP as an option seller}

It has been noted, for example in [9], that being a LP, if we consider the IL, is actually the same as being an option seller. In fact we know that we can replicate any given twice differentiable payoff $h(x)$ via the formula:

$$h(S_T)=h(S_0)+h'(S_0)(S_T-S_0)+\int_0^{S_0}h''(K)(K-S_T)^+\de K+\int_{S_0}^\infty h''(K)(S_T-K)^+\de K$$

In our case we have that our payoff function $h$ is the IL that is:

$$h(x)=\sqrt{\frac{x}{S_0}}-\frac{x}{2S_0}-\frac{1}{2}$$

clearly this function is $\mathcal{C}^\infty(\R^+)$ so that we can apply the replication formula. First we will compute the first and second derivative of the function:

$$\begin{dcases}
    h'(x)=\frac{1}{2\sqrt{xS_0}}-\frac{1}{2S_0} \\
    h''(x) = -\frac{1}{4x^{3/2}\sqrt{S_0}}<0
\end{dcases}$$

Thus substituting we obtain:

$$h(S_T)=-\int_0^{S_0}\frac{1}{4K^{3/2}\sqrt{S_0}}(K-S_T)^+\de K-\int_{S_0}^\infty \frac{1}{4K^{3/2}\sqrt{S_0}}(S_T-K)^+\de K$$

That is to say that we can replicate the IL selling an infinite strip of puts and calls of all strikes with maturity $T$.

\newpage

\section{LP pricing and Greeks}

On most DEXs like Uniswap, LPs can withdraw their liquidity at any time by redeeming their LP tokens. In this case, the price of the LP position is, as seen in formula (1.3.2), always equal to the value of the underlying assets plus fees:

$$P_t=V_0(\sqrt{r_t+1}+\phi t) = V_0\bigg(\sqrt{\frac{S_t}{S_0}}+\phi t\bigg)$$

Where $r_t$ is the return of asset $x$ relative to asset $y$ and $\phi$ is the expected APY.

\begingroup
\begin{center}
\includegraphics[scale=.34]{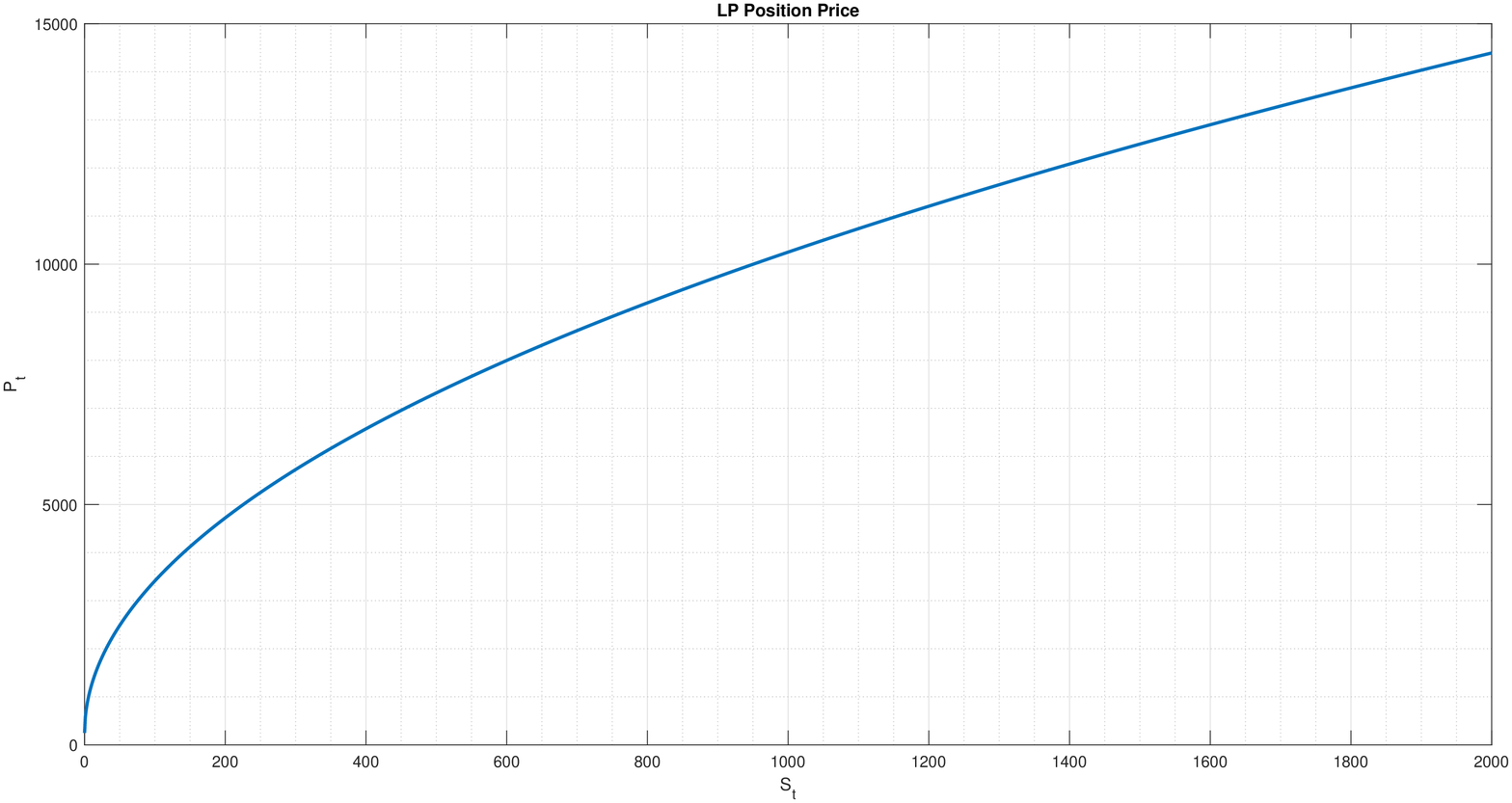}
\end{center}
\endgroup

As we will see below, unlocked liquidity has a positive delta, negative gamma and positive theta, while the vega exposure is zero. This is particularly important when options are used to hedge the position as suggested in [5] and [6]. Options do have a vega exposure, therefore an unlocked LP portfolio combined with a long options position always has a positive vega.

\subsection{Unlocked Liquidity Greeks}

\subsubsection{Delta}

Delta is defined as the partial derivative of the price of the position ($P_t$) with respect to the underlying price ($S_t$):

$$\Delta_{LP}\coloneqq\frac{\partial P_t}{\partial S_t} = \frac{V_0}{2\sqrt{S_0S_t}}$$

\begingroup
\begin{center}
\includegraphics[scale=.34]{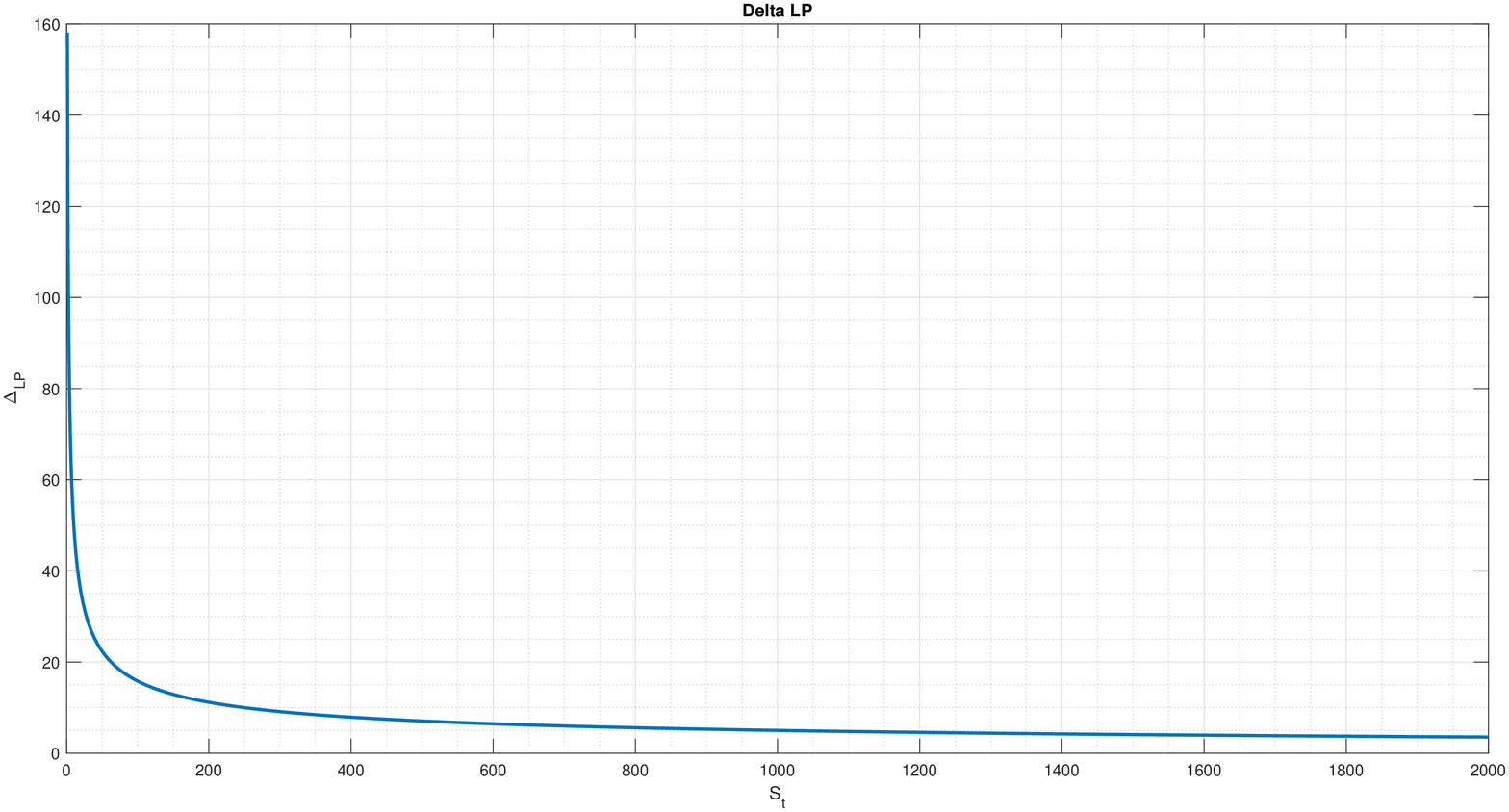}
\end{center}
\endgroup

\subsubsection{Delta 1\%}
Delta 1\% is defined as the change in price of the position when the underlying price changes by 1\%. So we will compute it as:

$$\Delta_{LP}^{1\%}(S_t)\coloneqq \Delta_{LP}(S_t)\cdot\frac{S_t}{100}=\frac{V_0}{2}\sqrt{\frac{S_t}{S_0}}\cdot 10^{-2}$$

\begingroup
\begin{center}
\includegraphics[scale=.34]{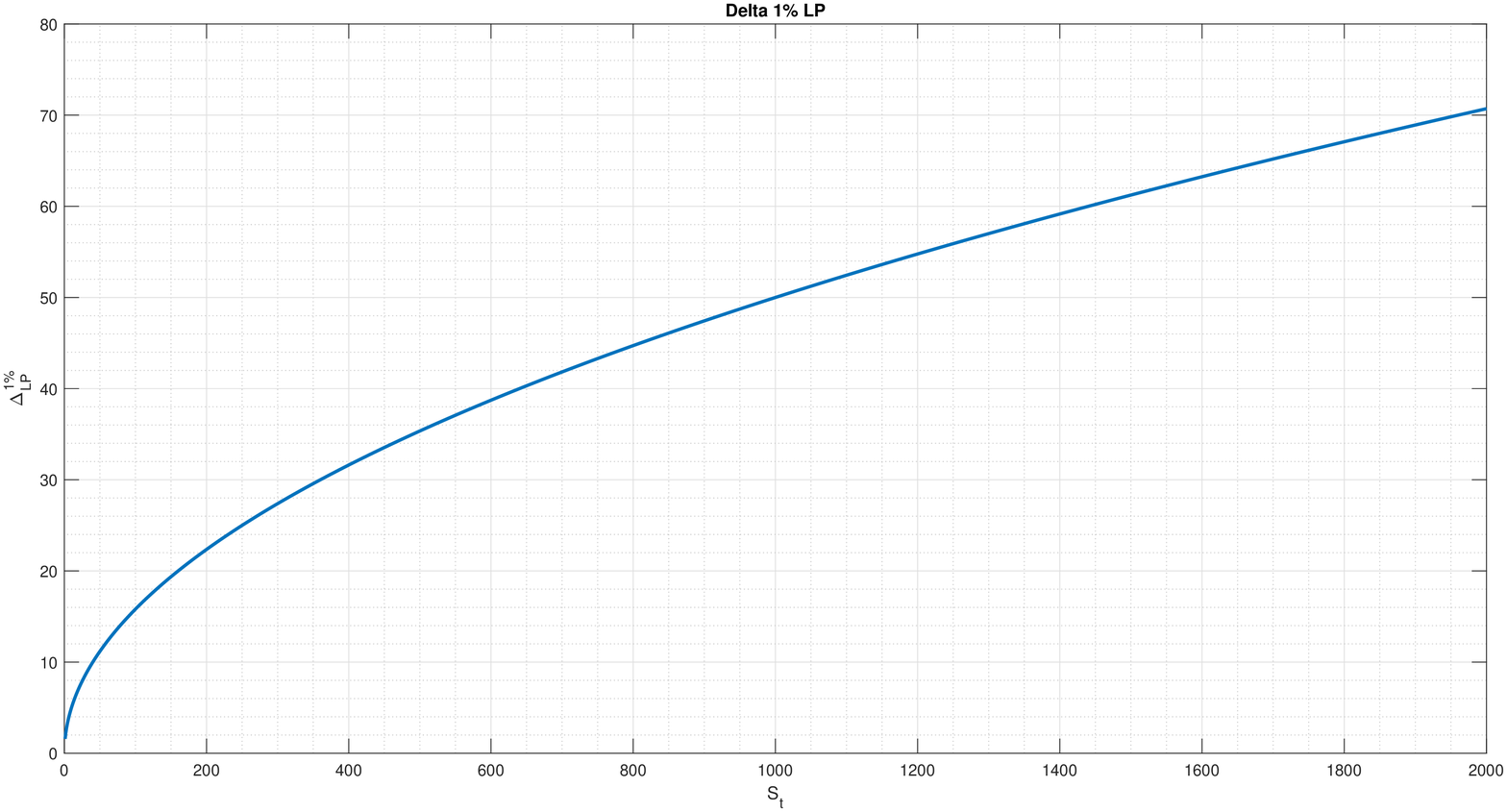}
\end{center}
\endgroup

\subsubsection{Gamma}

Gamma is defined as the second partial derivative of the price of the position ($P_t$) with respect to the underlying price ($S_t$). It can also be seen as the partial derivative of Delta with respect to the underlying price:

$$\Gamma_{LP}\coloneqq\frac{\partial^2 P_t}{\partial S_t^2}=\frac{\partial \Delta_{LP}}{\partial S_t}=-\frac{V_0}{4\sqrt{S_0}S_t^{3/2}}$$

We note that $\Gamma_{LP}<0$ and that $\Gamma_{LP}\to-\infty$ when $S_t\to0$.

\begingroup
\begin{center}
\includegraphics[scale=.34]{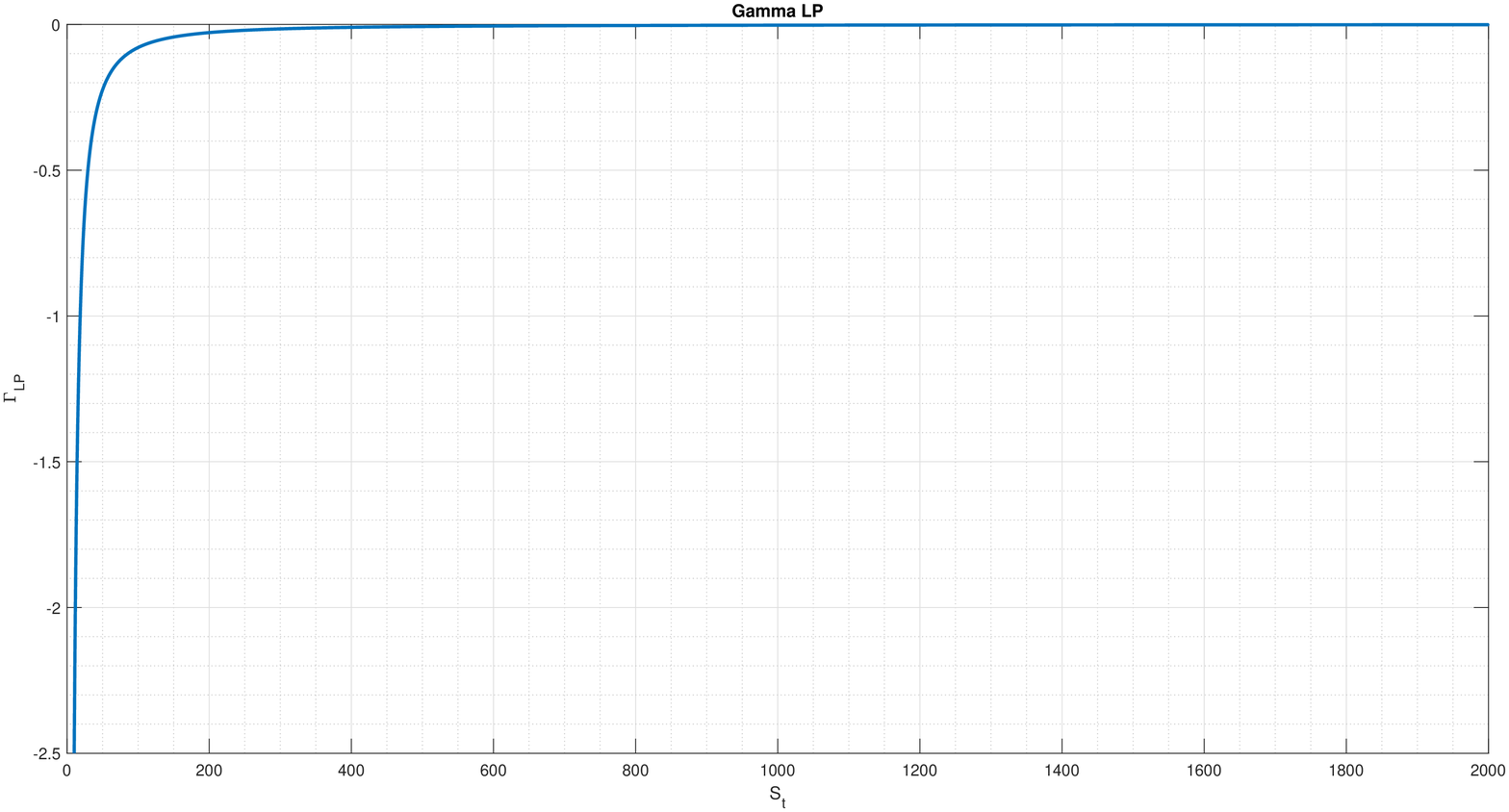}
\end{center}
\endgroup

\newpage

\subsubsection{Gamma 1\%}

Gamma 1\% is defined as the change of the Delta 1\% when the underlying price changes by 1\%. So we will compute it as:

$$\Gamma_{LP}^{1\%}(S_t)=\Gamma_{LP}(S_t)\cdot\bigg(\frac{S_t}{100}\bigg)^2=-\frac{V_0}{4}\sqrt{\frac{S_t}{S_0}}\cdot10^{-4}$$

\begingroup
\begin{center}
\includegraphics[scale=.34]{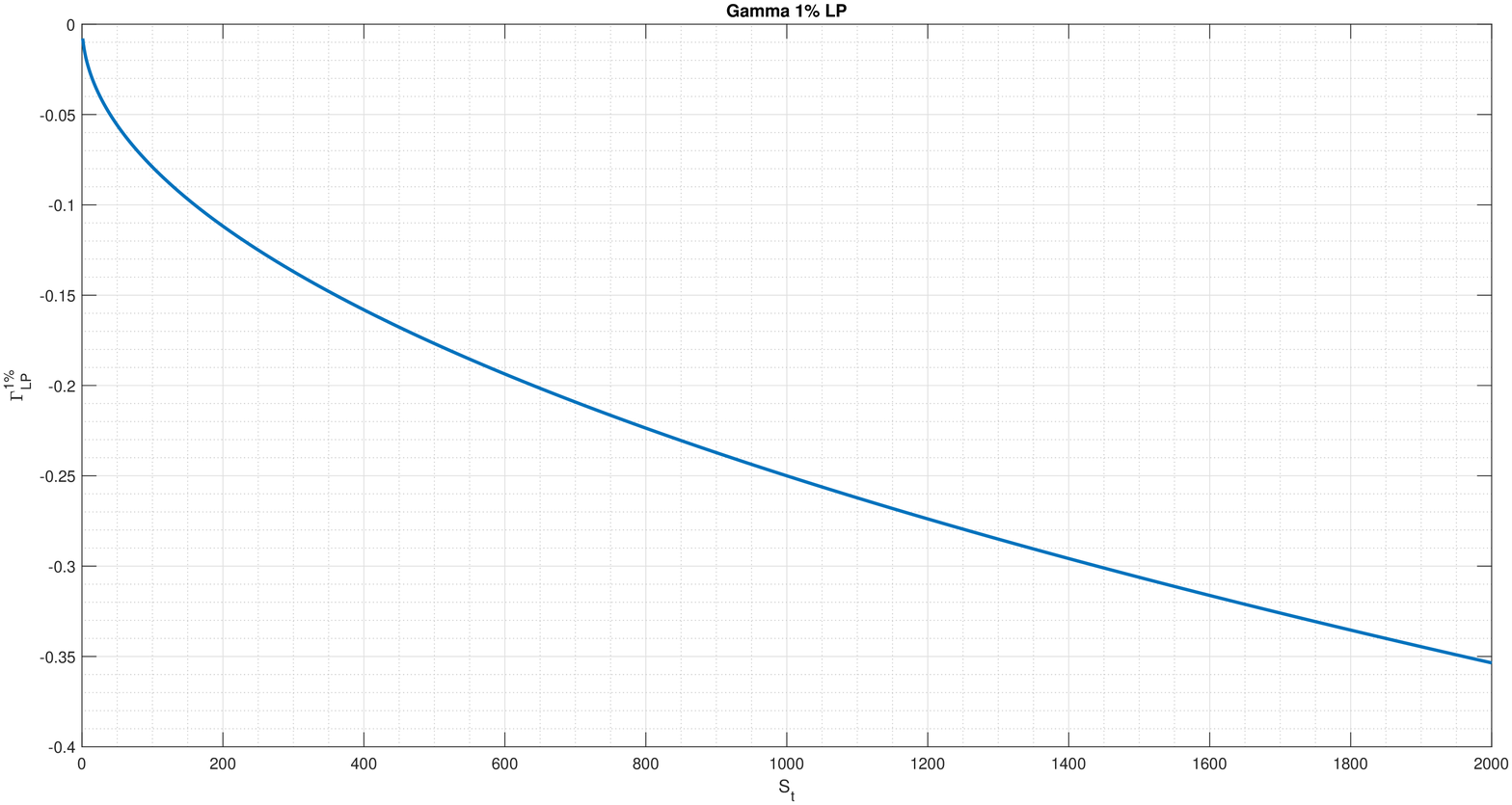}
\end{center}
\endgroup

\subsubsection{Vega}

Vega is defined as the partial derivative of the price of the position with respect to the volatility ($\sigma$):

$$\nu_{LP}\coloneqq\frac{\partial P_t}{\partial \sigma}=0$$

\subsubsection{Theta}

Theta is defined as the partial derivative of the price of the position ($P_t$) with respect to time ($t$):

$$\Theta_{LP}\coloneqq\frac{\partial P_t}{\partial t}=\phi \cdot V_0$$

\subsubsection{Rho}

Rho is defined as the partial derivative of the price of the position ($P_t$) with respect to the risk-free rate ($r_f$):

$$\rho_{LP}\coloneqq \frac{\partial P_t}{\partial r_f} = 0$$ 

\subsection{Locked Liquidity Analysis}

From here onwards, we assume that the Liquidity Provider has locked his liquidity until time $T$, perhaps to access liquidity mining rewards. When he does so, he will no longer be able to redeem the underlying assets until maturity. In this case, the fair price of his position may differ from the value of the underlying assets.

As we will see, while unlocked liquidity gives exposure only to delta, gamma and theta, locking the liquidity exposes the holder to additional vega and rho risks.

In order to calculate the fair price of the locked LP position we need to know:
\begin{itemize}
    \item the maturity $T$ when the liquidity gets unlocked (in years);
    \item the current time $t$ (in years);
    \item the remaining time $\tau=T-t$ (in years);
    \item the starting price for the underlying $S_0$;
    \item the initial capital invested $V_0$.
\end{itemize}

Similar to Black \& Scholes, we assume that the price processes are governed by the following SDEs:

$$\de B_t^x=r_xB_t^x\de t$$
$$\de B_t^y=r_yB_t^y\de t$$
$$\de S_t=\mu S_t\de t+\sigma S_t\de W_t$$

Where $B_t^x$ is the "bond" process relative to token $x$ with risk free rate $r_x$ (we can take for example the APY on lending token $x$ on Aave), $B_t^y$ is the "bond" process relative to token $y$ with risk free rate $r_y$ (analogously we can take the APY on lending token $y$ on Aave), $S_t$ is the market price of token $x$ in terms of token $y$ determined by the drift $\mu$, the volatility $\sigma$ and the Brownian Motion (BM) $W_t$. So this is to say that the price process is a Geometric Brownian Motion (GBM). It is known that we can find a probability $\mathcal{Q}$ (risk-free probability) in which we have that the price process follows the following SDE:

$$\de S_t=(r_x-r_y) S_t\de t + \sigma \de \Tilde{W}_t$$

Where we have that $\Tilde{W}_t$ is another BM given by the Ito formula:

$$\de \Tilde{W}_t=\de W_t-\frac{\mu-(r_x-r_y)}{\sigma}\de t$$

We can solve the price process SDE and defining $r_f\coloneqq r_x-r_y$ we obtain:

$$S_t=S_0\exp\bigg\{r_ft-\frac{1}{2}\sigma^2t+\sigma \Tilde{W}_t\bigg\}$$

We will assume that the liquidity pools are efficient, thanks to the presence of arbitrageurs, so that the pool price of token $x$ in terms of token $y$ is arbitrarily close to the market price, except at most in some instants, so that we can use the market price dynamics also for the pool price dynamics. From here onward we will denote everything in terms of token $y$. Now we can find the price of a locked liquidity position computing the discounted payoff under $\mathcal{Q}$:

$$P_t=\mathbb{E}_\mathcal{Q}[e^{-r_f\tau}V_{LP}(T)]=e^{-r_f\tau}\mathbb{E}_\mathcal{Q}\bigg[\frac{V_0}{\sqrt{S_0}}\sqrt{S_T}+\Phi(T)\bigg]=e^{-r_f\tau}\bigg(\frac{V_0}{\sqrt{S_0}}\mathbb{E}_\mathcal{Q}[\sqrt{S_T}]+\Phi(T)\bigg)$$

Thanks to the property of the GBM we have that:

$$\mathbb{E}_Q[\sqrt{S_T}]=\sqrt{S_t}\exp\bigg\{\frac{1}{2}r_f\tau-\frac{1}{8}\sigma^2 \tau\bigg\}\;\;\;\;\;(\star)$$

See Appendix A for the full derivation of $(\star)$. 
Substituting this into the pricing formula and doing some rearrangements we obtain:

$$P_t=V_0\bigg(\sqrt{\frac{S_t}{S_0}}\exp\bigg\{-\tau\bigg(\frac{r_f}{2}+\frac{\sigma^2}{8}\bigg)\bigg\}+\phi Te^{-r_f\tau}\bigg)$$

This is the fair value of the LP position locked until time $T$, when the expected volatility on the pair of tokens is $\sigma$ and $\phi$ is the expected APY of the pool.

\begingroup
\begin{center}
\includegraphics[scale=.38]{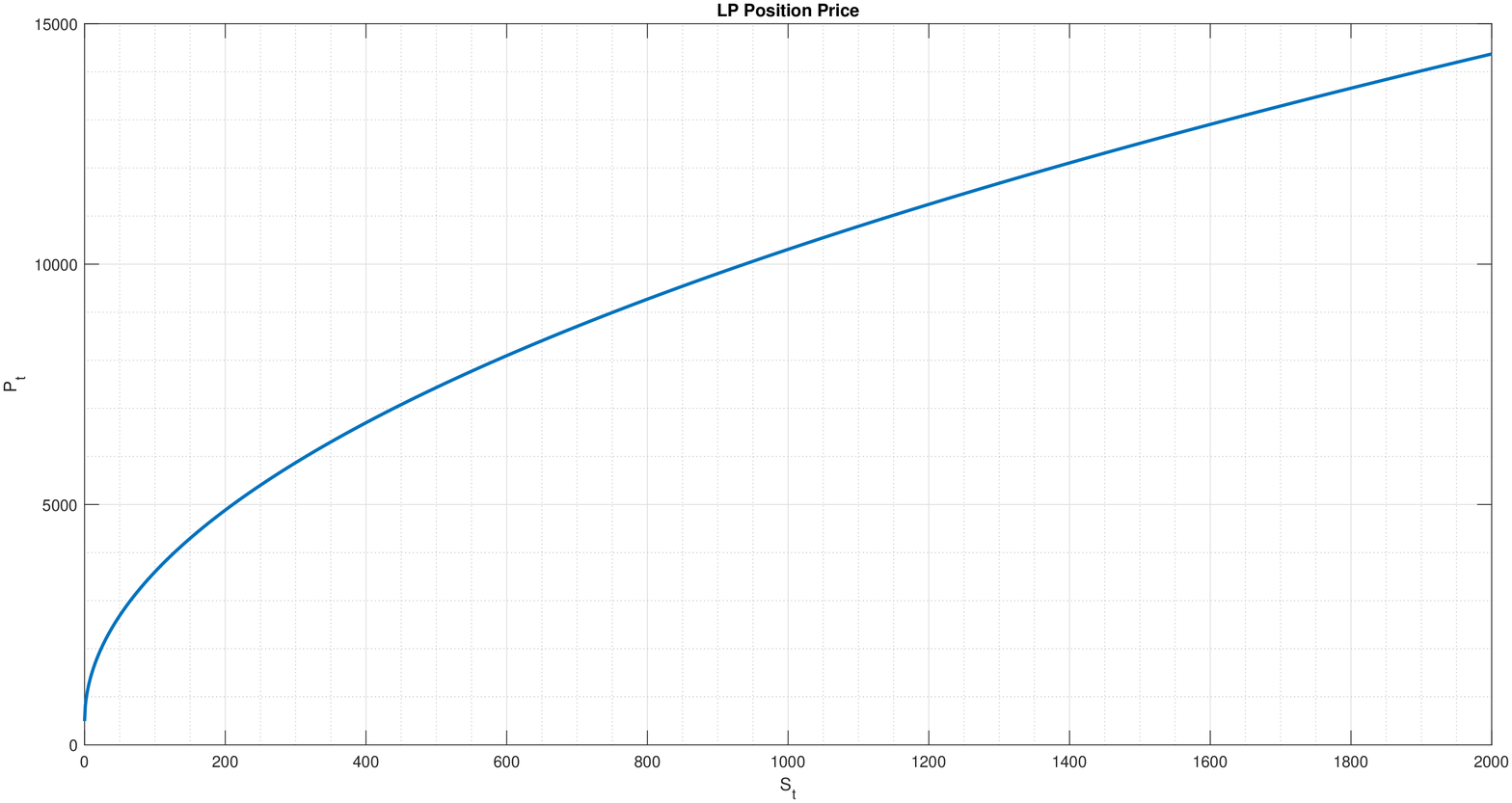}
\end{center}
\endgroup

Plot obtained with the following data:
\begin{itemize}
    \item $V_0=10000$, $S_0 = 1000$;
    \item $r_f=3\%$, $\sigma=70\%$, $\phi = 10\%$;
    \item $T = 0.5$, $\tau = 0.25$.
\end{itemize}
All the plots in this section are obtained using the same data.

\subsection{Greeks}

\subsubsection{Delta}

$$\Delta_{LP}=\frac{V_0}{2\sqrt{S_0S_t}}\exp\bigg\{-\tau\bigg(\frac{r_f}{2}+\frac{\sigma^2}{8}\bigg)\bigg\}$$

We note that $\Delta_{LP}>0$.

\begingroup
\begin{center}
\includegraphics[scale=.38]{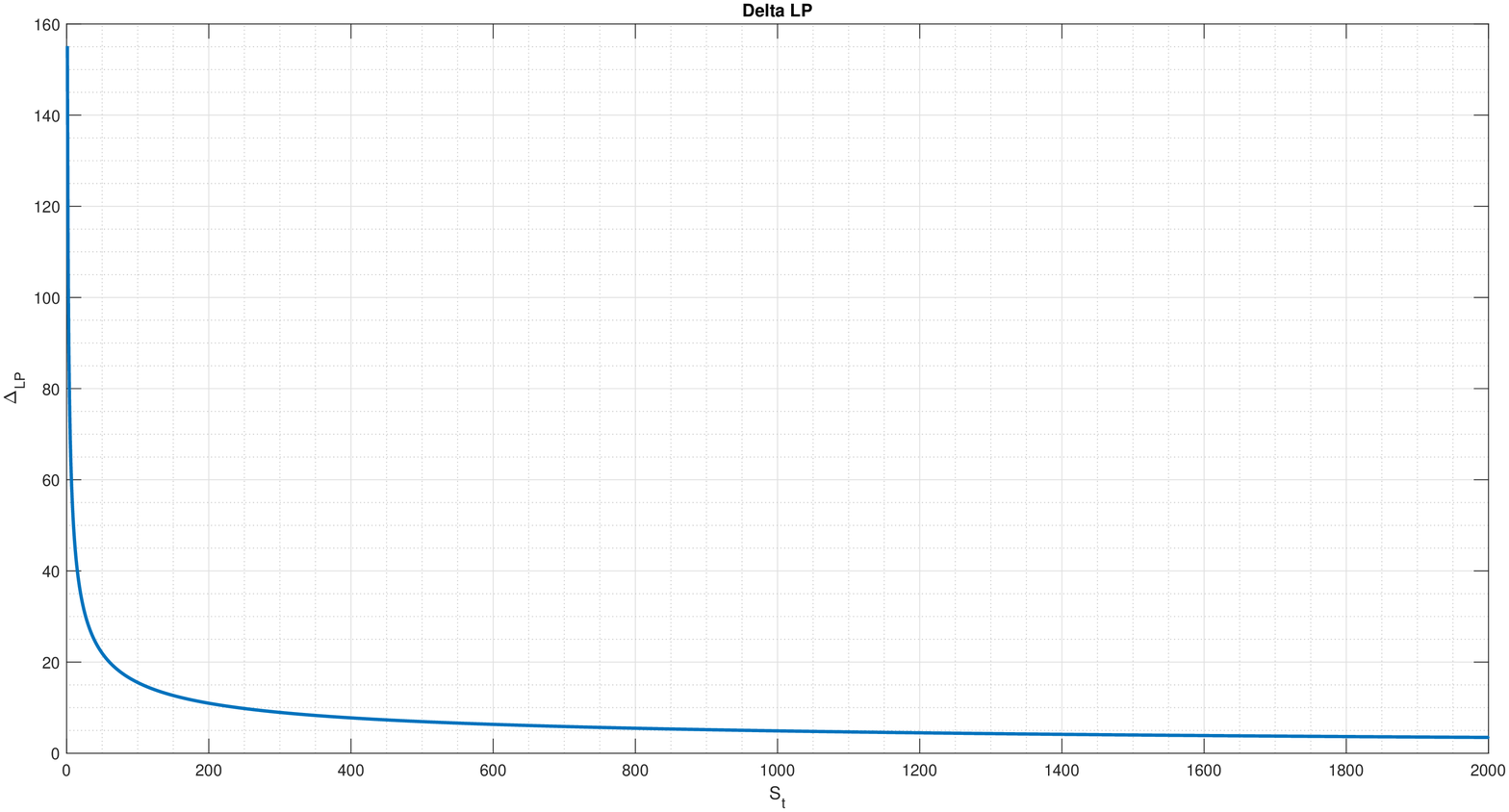}
\end{center}
\endgroup

We can see that the delta of a liquidity provider increases when the underlying price S drops and vice versa.

\subsubsection{Delta 1\%}

$$\Delta_{LP}^{1\%}(S_t) =\frac{V_0}{2}\sqrt{\frac{S_t}{S_0}}\exp\bigg\{-\tau\bigg(\frac{r_f}{2}+\frac{\sigma^2}{8}\bigg)\bigg\}\cdot10^{-2}$$

\begingroup
\begin{center}
\includegraphics[scale=.38]{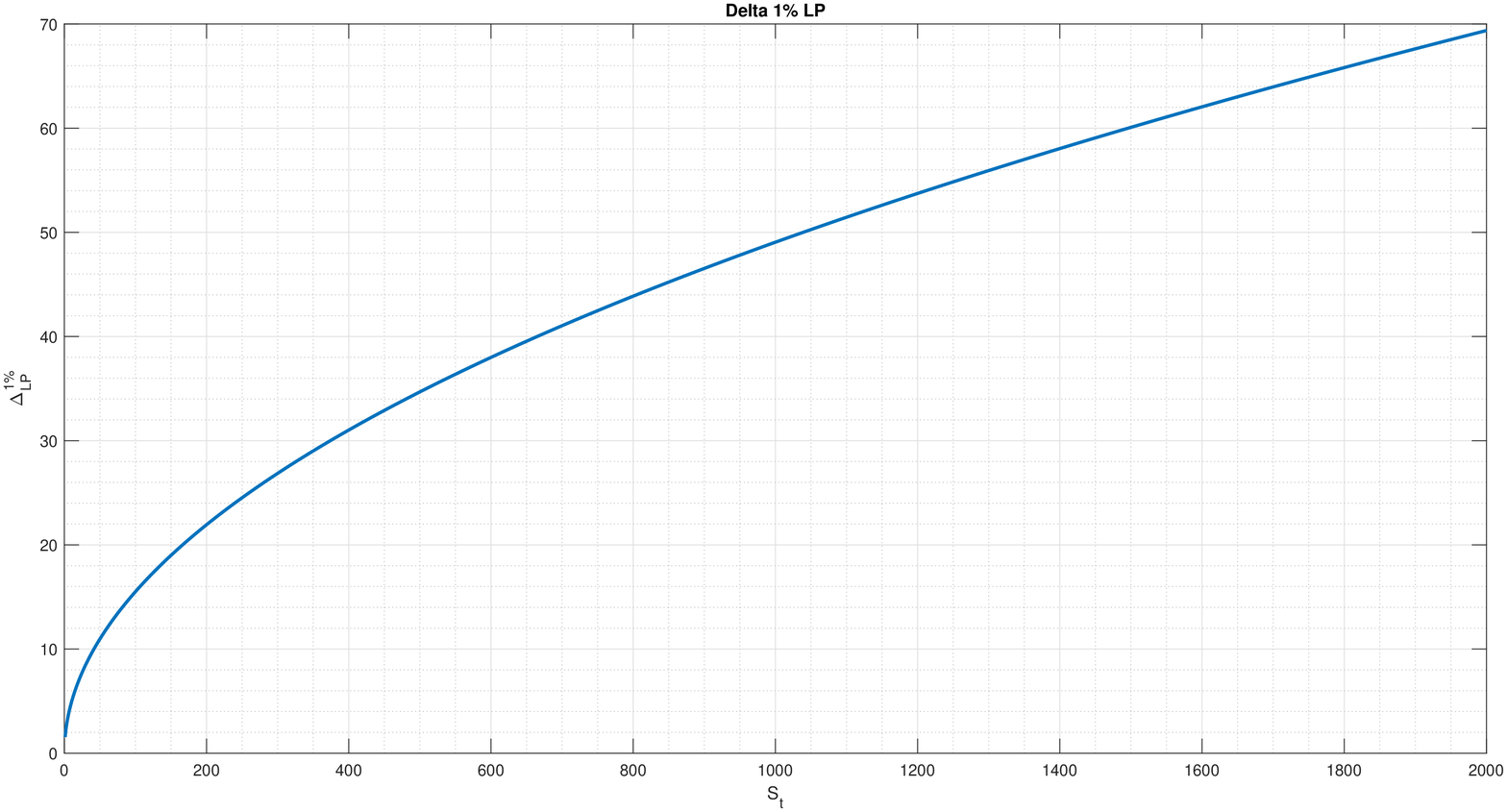}
\end{center}
\endgroup

\subsubsection{Gamma}

$$\Gamma_{LP}=-\frac{V_0}{4\sqrt{S_0}S_t^{3/2}}\exp\bigg\{-\tau\bigg(\frac{r_f}{2}+\frac{\sigma^2}{8}\bigg)\bigg\}$$

We note that $\Gamma_{LP}<0$ and that $\Gamma_{LP}\to-\infty$ when $S_t\to0$.

\begingroup
\begin{center}
\includegraphics[scale=.38]{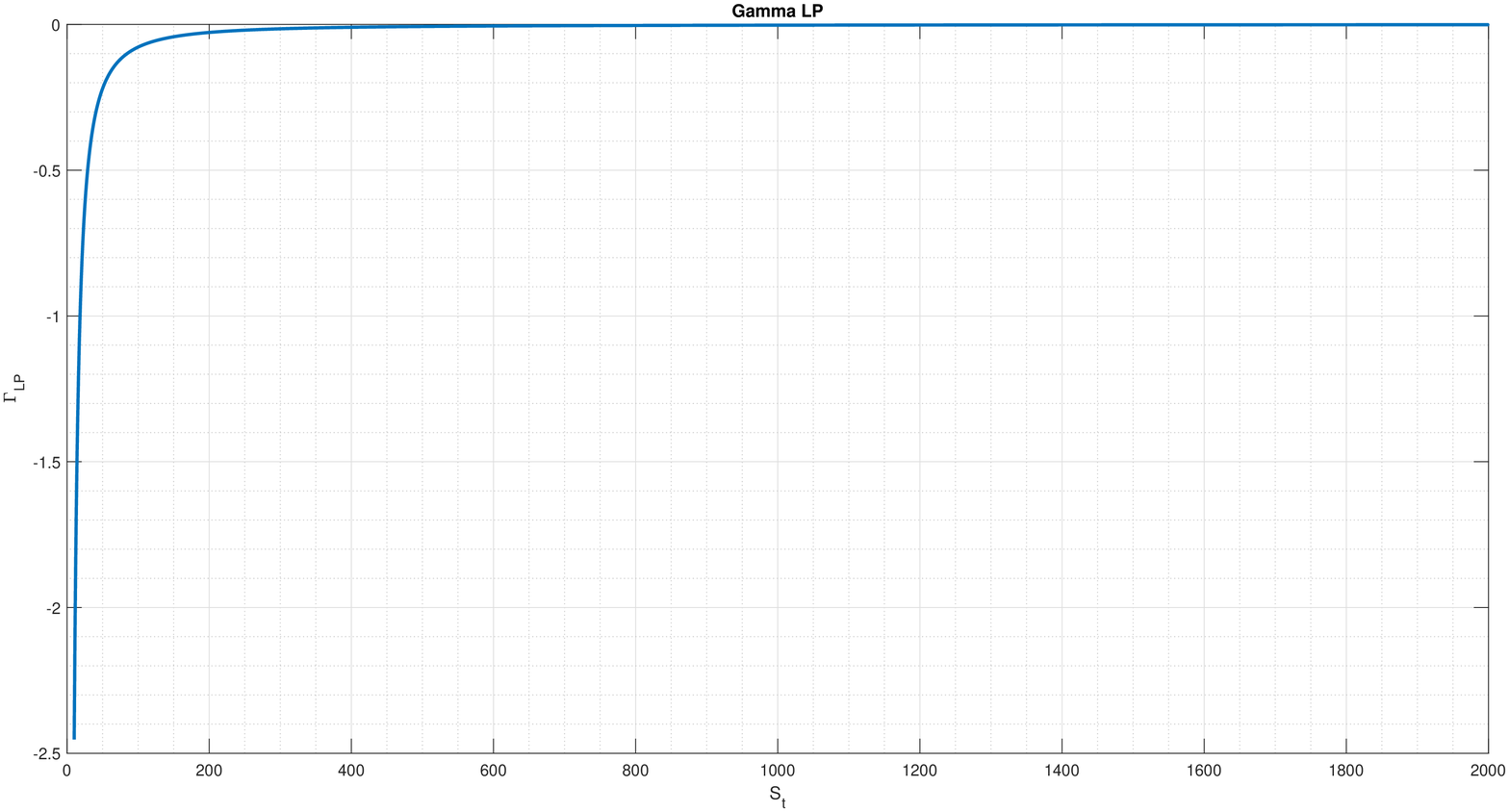}
\end{center}
\endgroup

\subsubsection{Gamma 1\%}

$$\Gamma_{LP}^{1\%}(S_t)=-\frac{V_0}{4}\sqrt{\frac{S_t}{S_0}}\exp\bigg\{-\tau\bigg(\frac{r_f}{2}+\frac{\sigma^2}{8}\bigg)\bigg\}\cdot 10^{-4}$$

\begingroup
\begin{center}
\includegraphics[scale=.38]{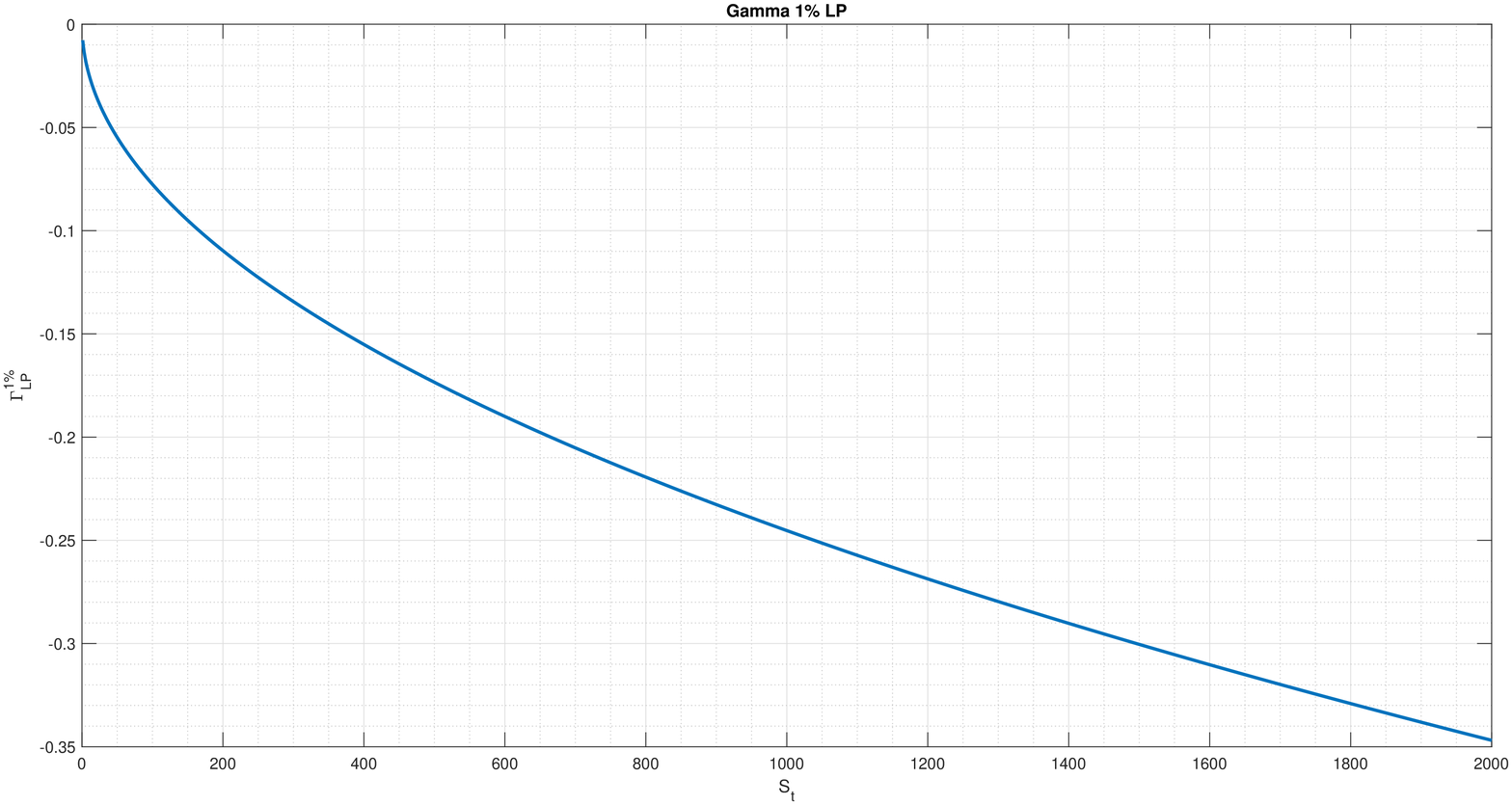}
\end{center}
\endgroup

\subsubsection{Vega}

$$\nu_{LP}=-V_0\frac{\sigma\tau}{4}\sqrt{\frac{S_t}{S_0}}\exp\bigg\{-\tau\bigg(\frac{r_f}{2}+\frac{\sigma^2}{8}\bigg)\bigg\}$$

We note that $\nu_{LP}<0$.

\begingroup
\begin{center}
\includegraphics[scale=.38]{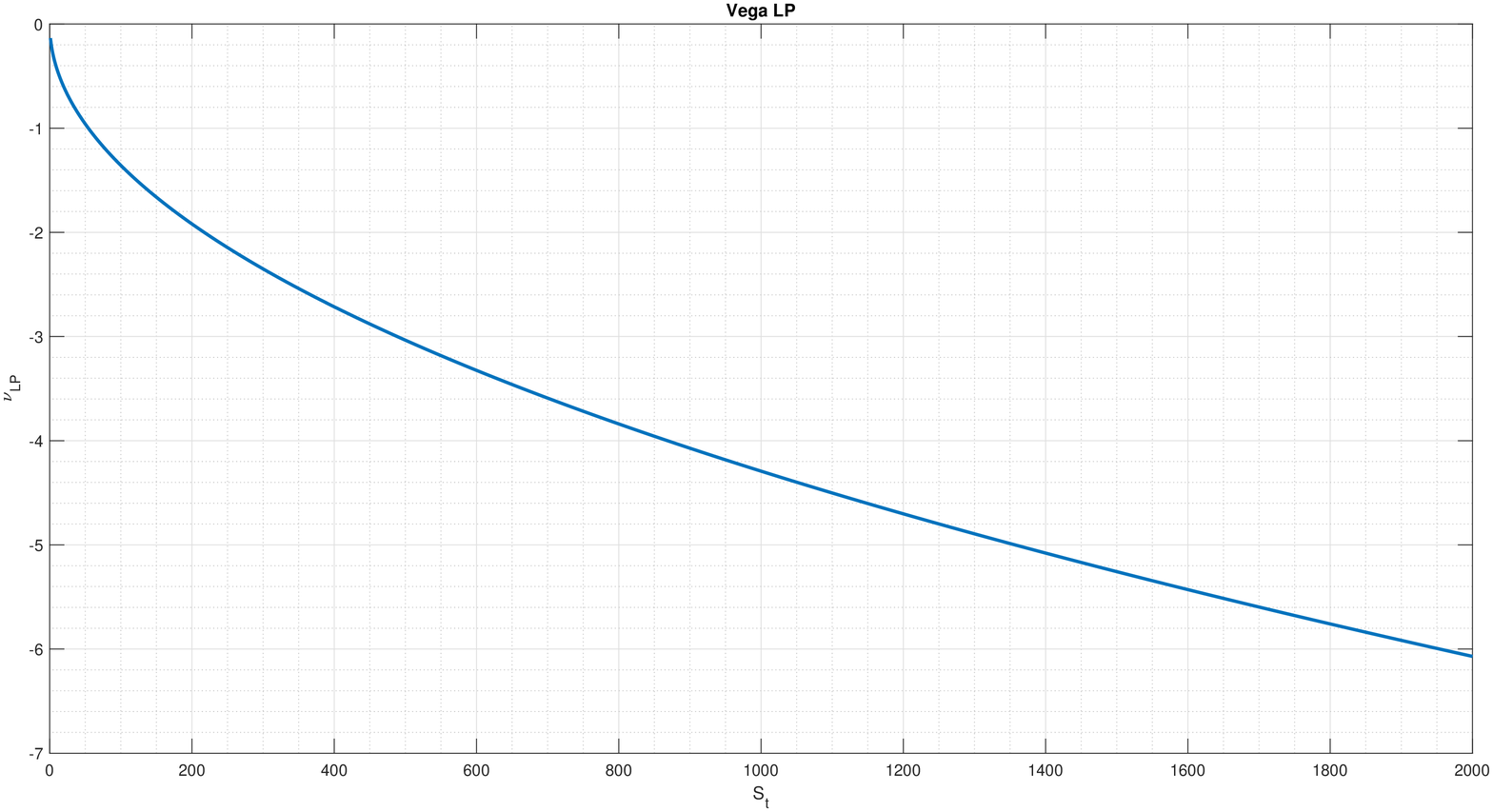}
\end{center}
\endgroup

In the figure we have plotted the Vega corresponding to a 1\% change in volatility $\sigma$, that is $\nu/100$.

\subsubsection{Theta}

$$\Theta_{LP}=V_0\bigg(\sqrt{\frac{S_t}{S_0}}\bigg(\frac{r_f}{2}+\frac{\sigma^2}{8}\bigg)\exp\bigg\{-\tau\bigg(\frac{r_f}{2}+\frac{\sigma^2}{8}\bigg)\bigg\}+r_f\phi Te^{-r_f\tau}\bigg)$$

\begingroup
\begin{center}
\includegraphics[scale=.38]{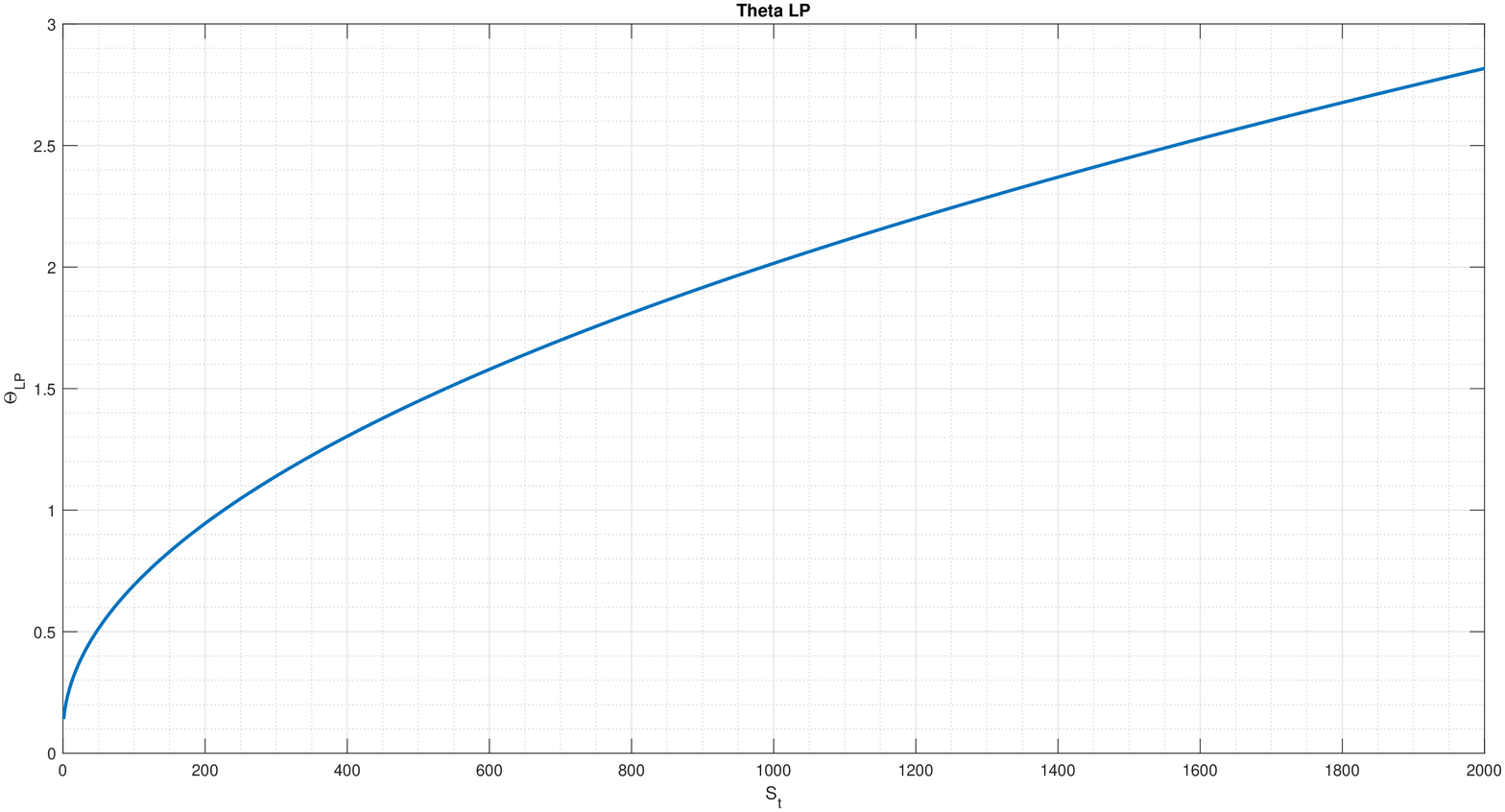}
\end{center}
\endgroup

In the figure we have plotted the  daily Theta, that is $\Theta/365$.

\subsubsection{Rho}

Even if we have a model with two risk free rates we note that the price depends only on their difference $r_f$ so we can study only the sensibility with respect to $r_f$.

$$\rho_{LP}\coloneqq\frac{\partial P_t}{\partial r_f}=-V_0\bigg(\frac{\tau}{2}\sqrt{\frac{S_t}{S_0}}\exp\bigg\{-\tau\bigg(\frac{r_f}{2}+\frac{\sigma^2}{8}\bigg)\bigg\}+\tau\phi T e^{-r_f\tau}\bigg)$$

\begingroup
\begin{center}
\includegraphics[scale=.38]{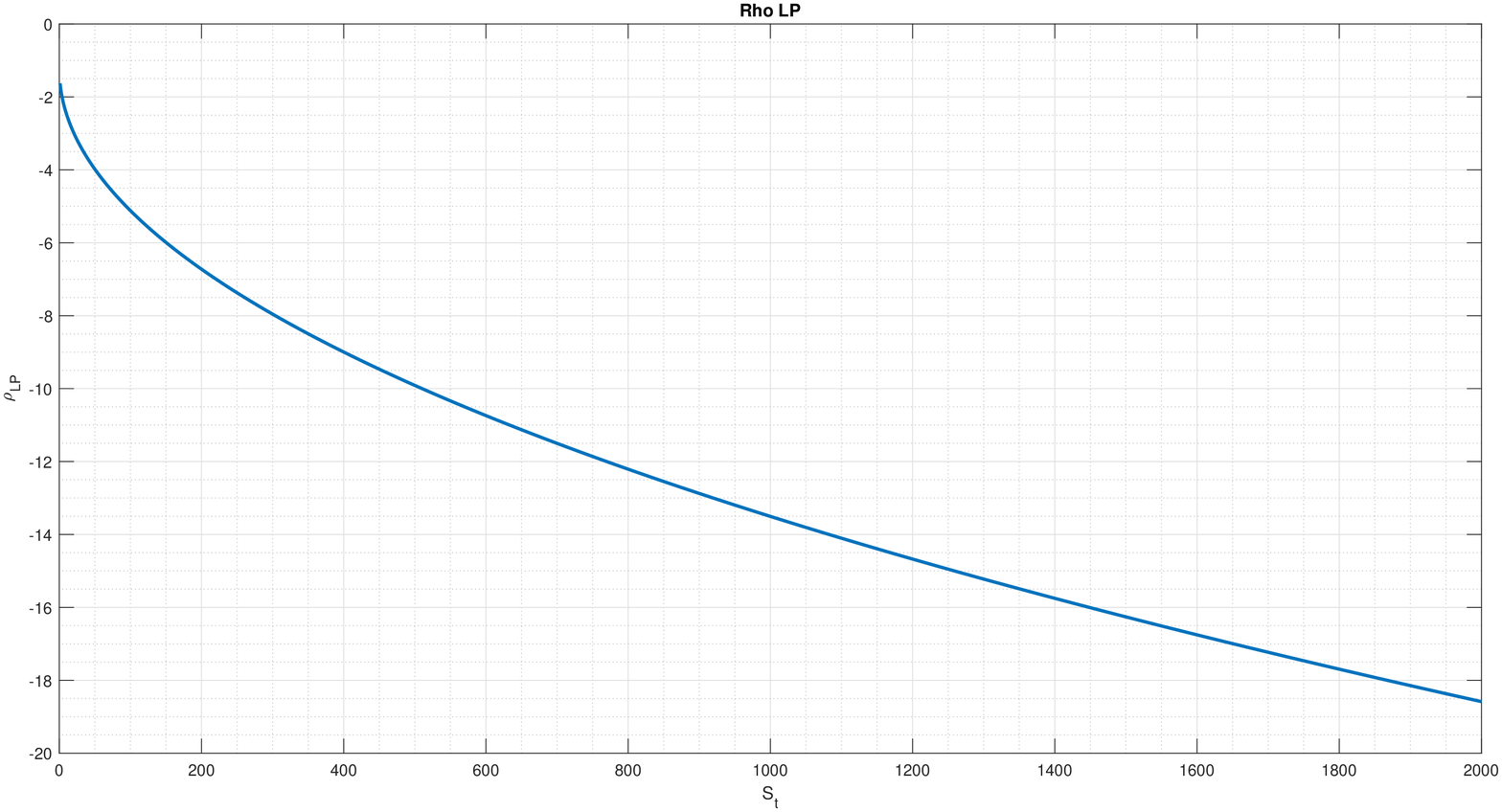}
\end{center}
\endgroup

In the figure we have plotted the Rho corresponding to a 1\% change of $r_f$, that is $\rho/100$.

\newpage

\section{Impermanent Gain}

As detailed in the previous section, a LP is exposed to many risks. While it is relatively easy to hedge the Delta, for example shorting futures on the underlying $S$, we ask ourselves if we can structure a product that hedges all the other greeks, in particular Vega, Gamma and Theta. We call this product Impermanent Gain (IG).

IG's unit payoff is defined as the opposite of IL:

$$IG(r)=\frac{V_H-V_{LP}}{V_0}=1+\frac{r}{2}-\sqrt{r+1}$$

The IG is a product that has some similarities with European options:

\begin{itemize}
    \item it has a maturity $T$;
    \item it has a strike $K$ that is the starting price from which the IG is computed:
    $$r_t=\frac{S_t}{K}-1$$
    where $r_t$ is used to compute the IG at time $t$.
\end{itemize}

IG payoff with maturity $T$ can be coded into a smart contract making it a DeFi-native product.

\subsection{Pricing}

We have seen in \textbf{Section 2.1} that we can replicate the IL selling an infinite strip of puts and calls and so we can also replicate the IG buying the same portfolio, thus we could price the IG position finding the cost of the replicating portfolio. Instead of doing that we will use a different approach giving a dynamic for the price process of token $x$, in terms of token $y$, and using as the price for the IG position the discounted payoff. We will assume the same dynamics and conditions described in \textbf{Section 3.2} for the market and liquidity pool price processes.
Given the following data: 
\begin{itemize}
    \item maturity $T$ (in years);
    \item current time $t$ (in years);
    \item time to expiry $\tau=T-t$ (in years);
    \item strike $K=S_0$, that is the starting price of the token;
    \item initial capital $V_0$;
\end{itemize}
we can calculate the price of the IG strategy at time $t$ as the discounted payoff under the risk-free measure $\mathcal{Q}$:

$$P_t=\mathbb{E}_\mathcal{Q}[e^{-r_f\tau} \cdot V_0\cdot IG(S_T)]=\mathbb{E}_\mathcal{Q}\bigg[e^{-r_f\tau}V_0\bigg(\frac{1}{2}+\frac{S_T}{2K}-\sqrt{\frac{S_T}{K}}\bigg)\bigg]$$

$$P_t=e^{-r_f\tau}V_0\bigg(\frac{1}{2}+\frac{1}{2K}\mathbb{E}_\mathcal{Q}[S_T]-\frac{1}{\sqrt{K}}\mathbb{E}_\mathcal{Q}[\sqrt{S_T}]\bigg)$$

Remembering the properties of the GBM we have:

$$\mathbb{E}_\mathcal{Q}[S_T]=S_te^{r_f\tau}\;\;\;\;\;(\bullet)$$

$$\mathbb{E}_\mathcal{Q}[\sqrt{S_T}]=\sqrt{S_t}\exp\bigg\{\frac{1}{2}r_f\tau-\frac{1}{8}\sigma^2 \tau\bigg\}$$

See Appendix B for the full derivation of $(\bullet)$. Substituting into the price formula and doing some rearrangements we obtain:

$$P_t=V_0\bigg(\frac{1}{2}e^{-r_f\tau}+\frac{S_t}{2K}-\sqrt{\frac{S_t}{K}}\exp\bigg\{-\bigg(\frac{r_f}{2}+\frac{\sigma^2}{8}\bigg)\tau\bigg\}\bigg)$$

\begingroup
\begin{center}
\includegraphics[scale=.38]{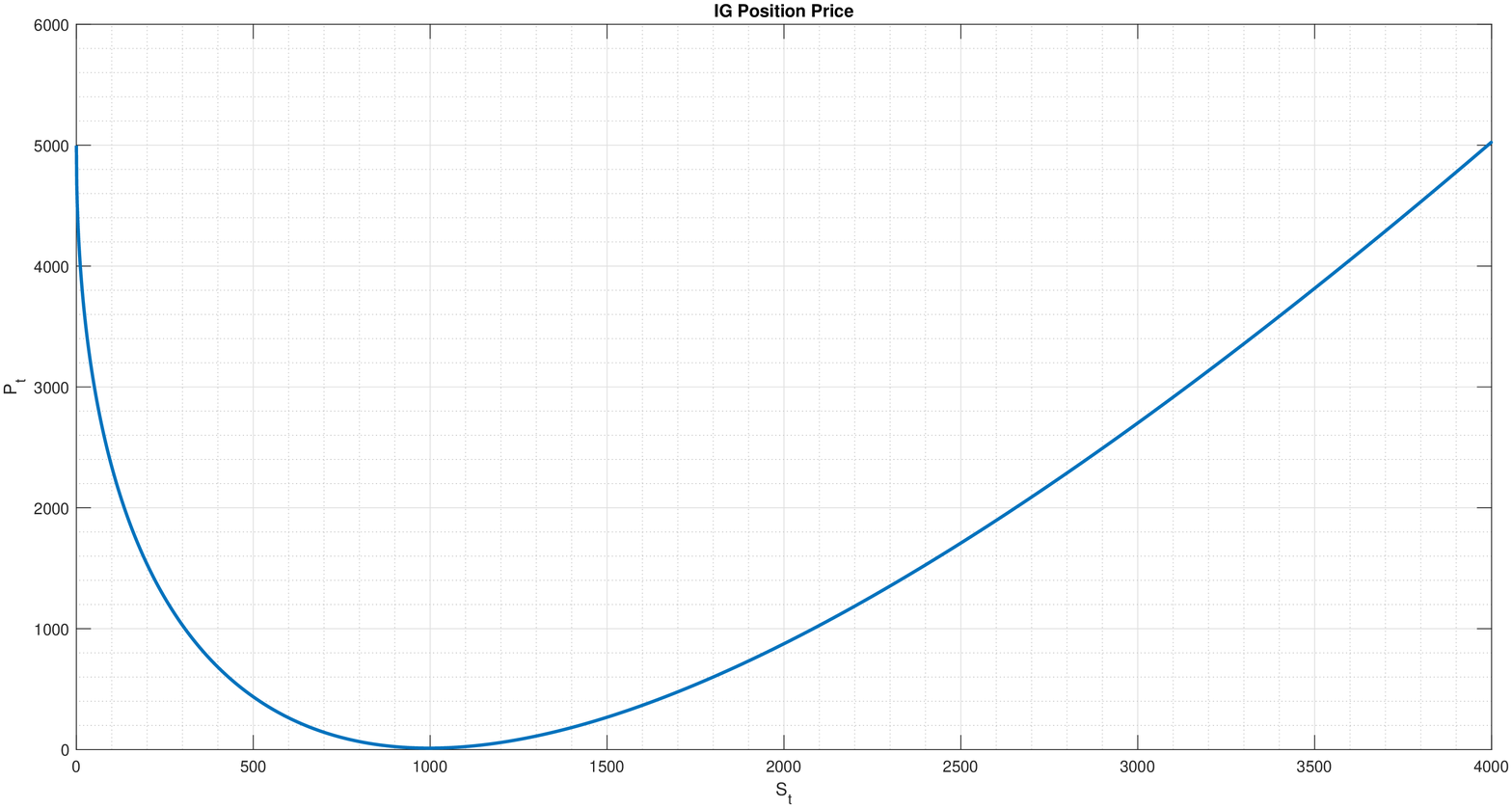}
\end{center}
\endgroup

Plot obtained with the following data:
\begin{itemize}
    \item $V_0=10000$, $K = 1000$;
    \item $r_f=3\%$, $\sigma=70\%$;
    \item $\tau = \frac{7}{365}$ (seven days).
\end{itemize}
All the plots in this section are obtained using the same data.

\subsection{Greeks}

\subsubsection{Delta}

$$\Delta_{IG}=V_0\bigg(\frac{1}{2K}-\frac{1}{2\sqrt{KS_t}}\exp\bigg\{-\bigg(\frac{r_f}{2}+\frac{\sigma^2}{8}\bigg)\tau\bigg\}\bigg)$$

We note that the $\Delta_{IG}(K)>0$.

\begingroup
\begin{center}
\includegraphics[scale=.38]{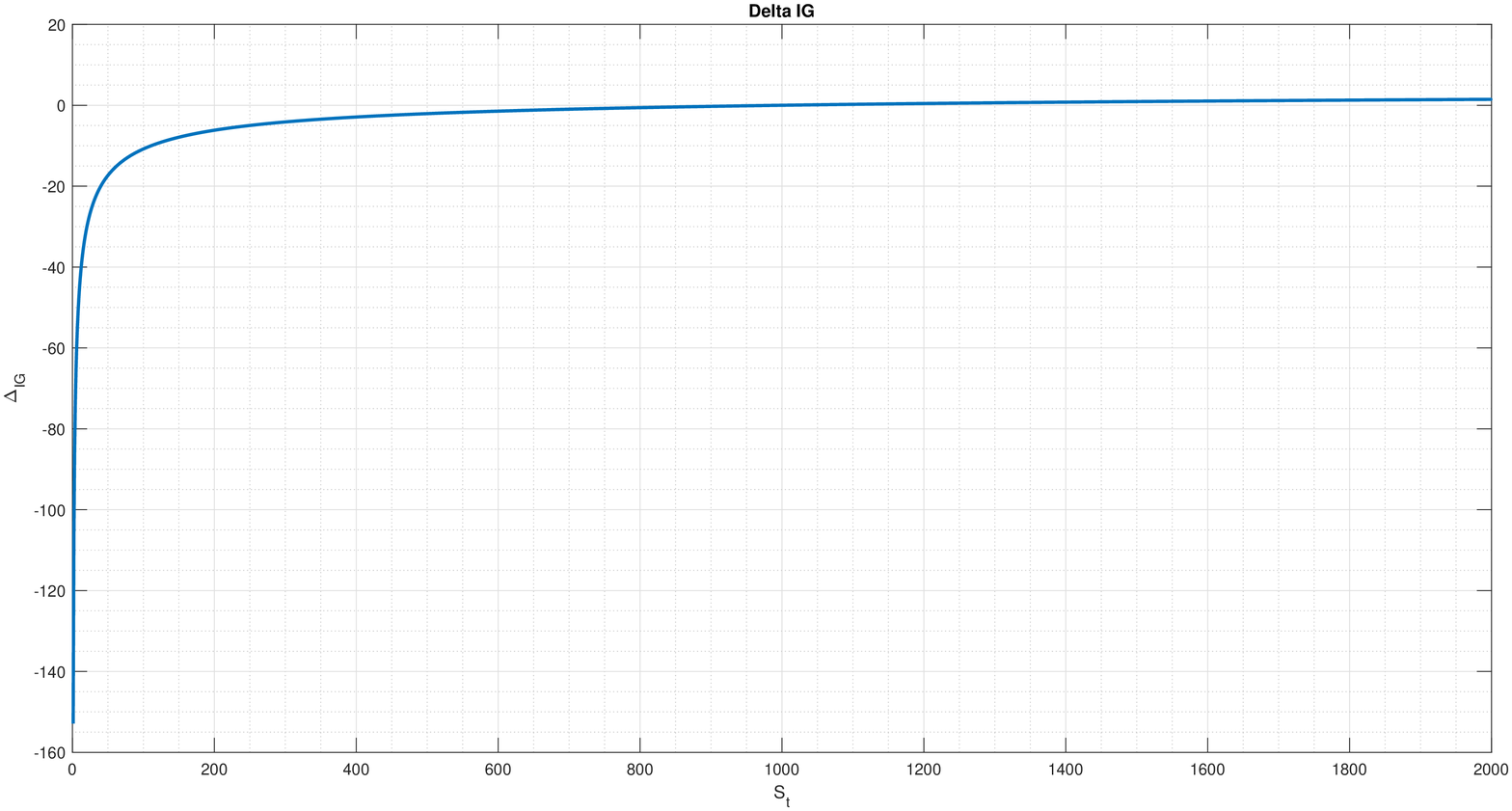}
\end{center}
\endgroup

\subsubsection{Delta 1\%}

$$\Delta_{IG}^{1\%}(S_t)=V_0\bigg(\frac{S_t}{2K}-\frac{1}{2}\sqrt{\frac{S_t}{K}}\exp\bigg\{-\bigg(\frac{r_f}{2}+\frac{\sigma^2}{8}\bigg)\tau\bigg\}\bigg)\cdot10^{-2}$$

\begingroup
\begin{center}
\includegraphics[scale=.38]{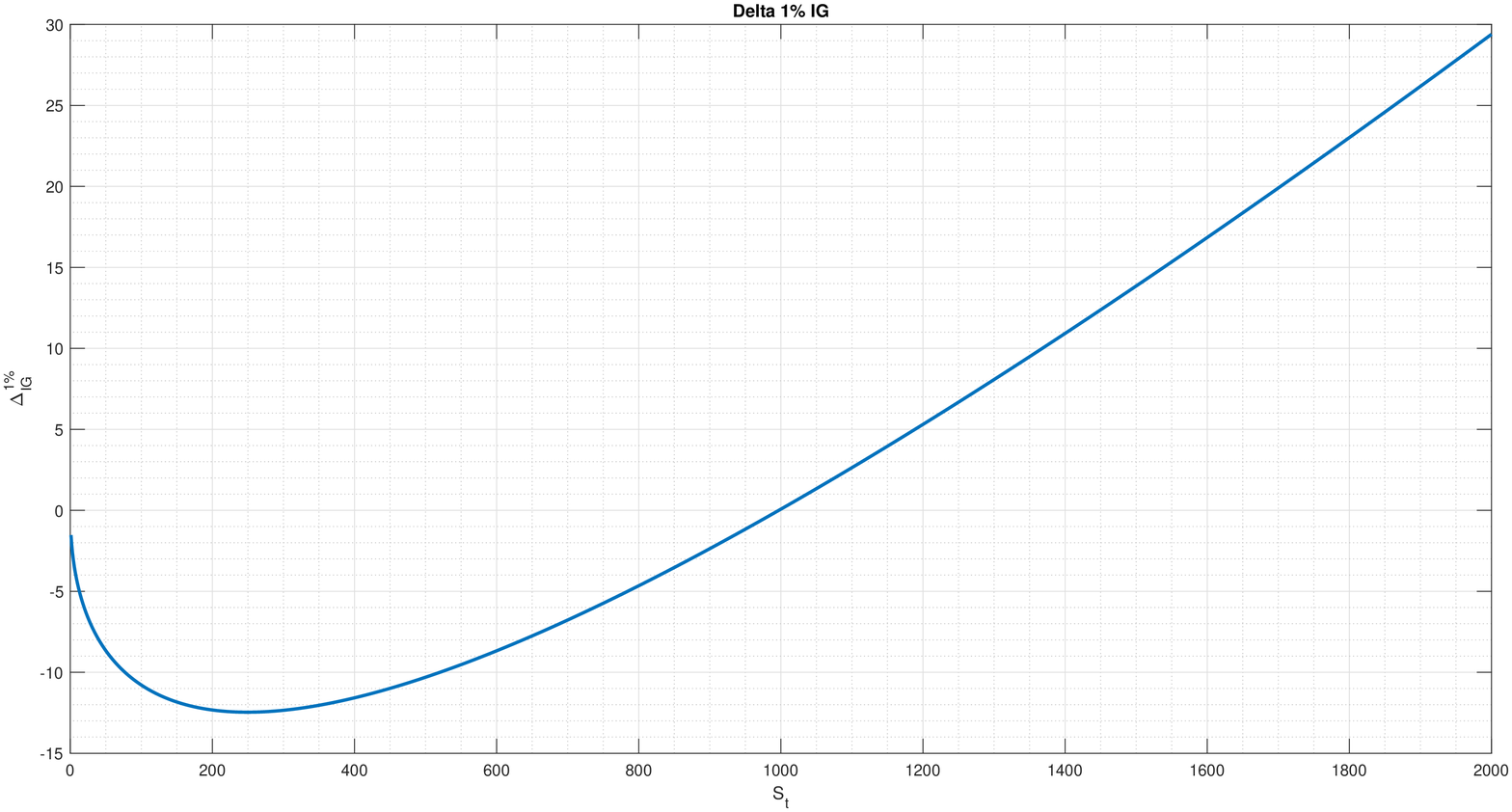}
\end{center}
\endgroup

\subsubsection{Gamma}

$$\Gamma_{IG}=\frac{\partial^2 P_t}{\partial S_t^2}=\frac{V_0}{4\sqrt{K}S_t^{3/2}}\exp\bigg\{-\bigg(\frac{r_f}{2}+\frac{\sigma^2}{8}\bigg)\tau\bigg\}$$

We note that $\Gamma_{IG}>0$ and that $\Gamma_{LP}\to+\infty$ when $S_t\to0$.

\begingroup
\begin{center}
\includegraphics[scale=.38]{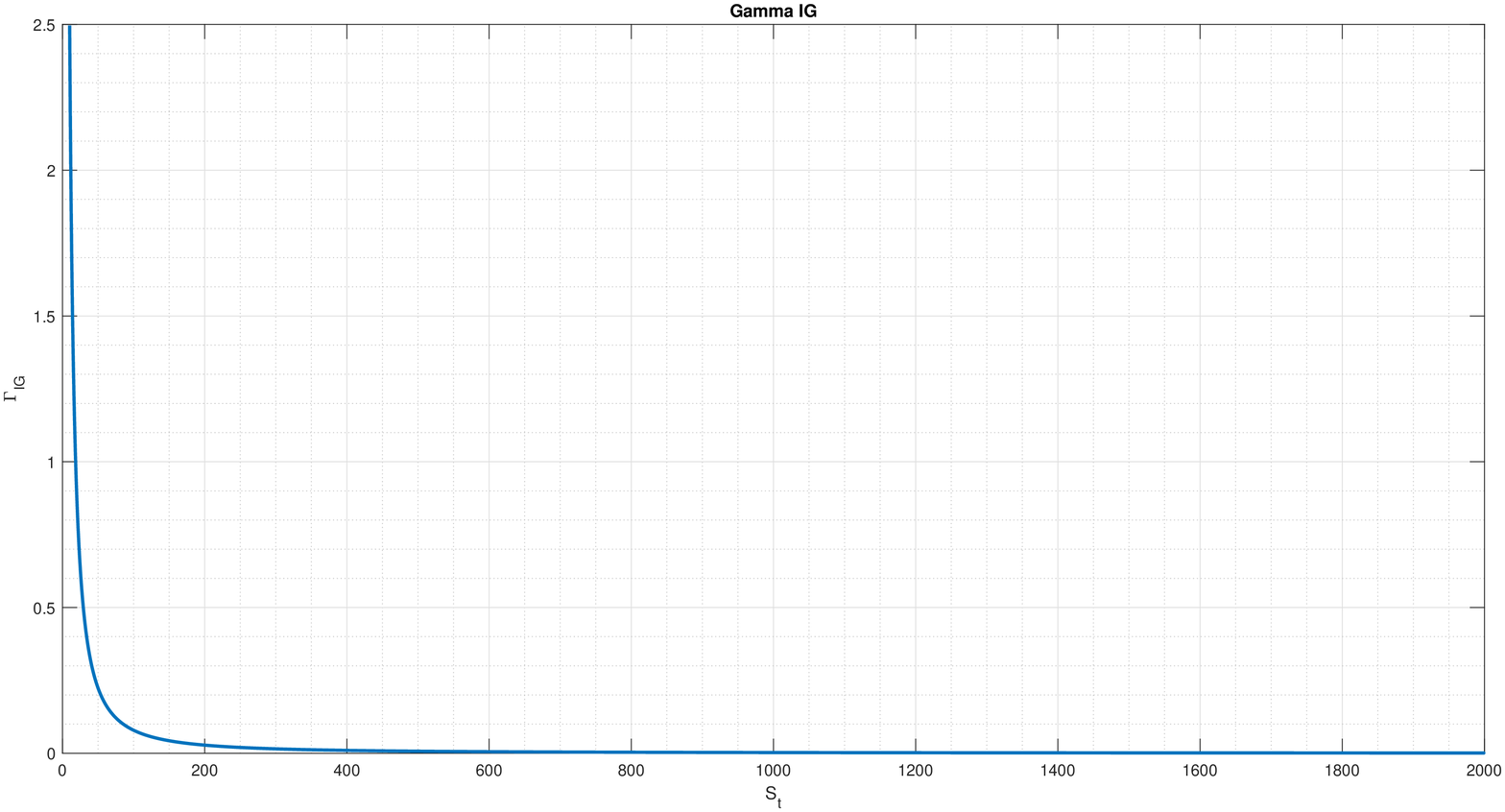}
\end{center}
\endgroup

\subsubsection{Gamma 1\%}

$$\Gamma_{IG}^{1\%}(S_t)=\frac{V_0}{4}\sqrt{\frac{S_t}{K}}\exp\bigg\{-\bigg(\frac{r_f}{2}+\frac{\sigma^2}{8}\bigg)\tau\bigg\}\cdot 10^{-4}$$

\begingroup
\begin{center}
\includegraphics[scale=.38]{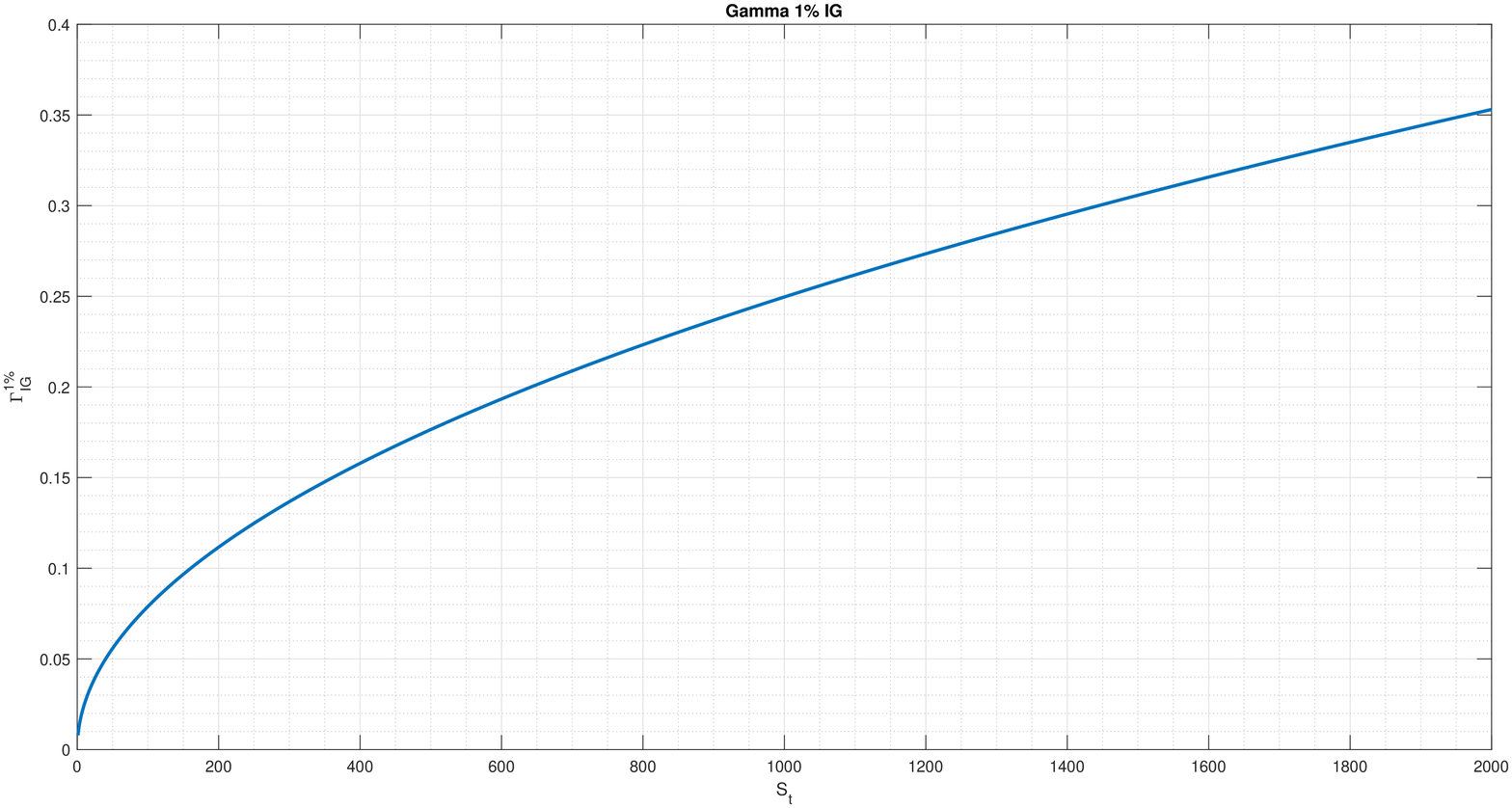}
\end{center}
\endgroup

\subsubsection{Vega}

$$\nu_{IG}=V_0\frac{\sigma \tau}{4}\sqrt{\frac{S_t}{K}}\exp\bigg\{-\bigg(\frac{r_f}{2}+\frac{\sigma^2}{8}\bigg)\tau\bigg\}$$

\begingroup
\begin{center}
\includegraphics[scale=.38]{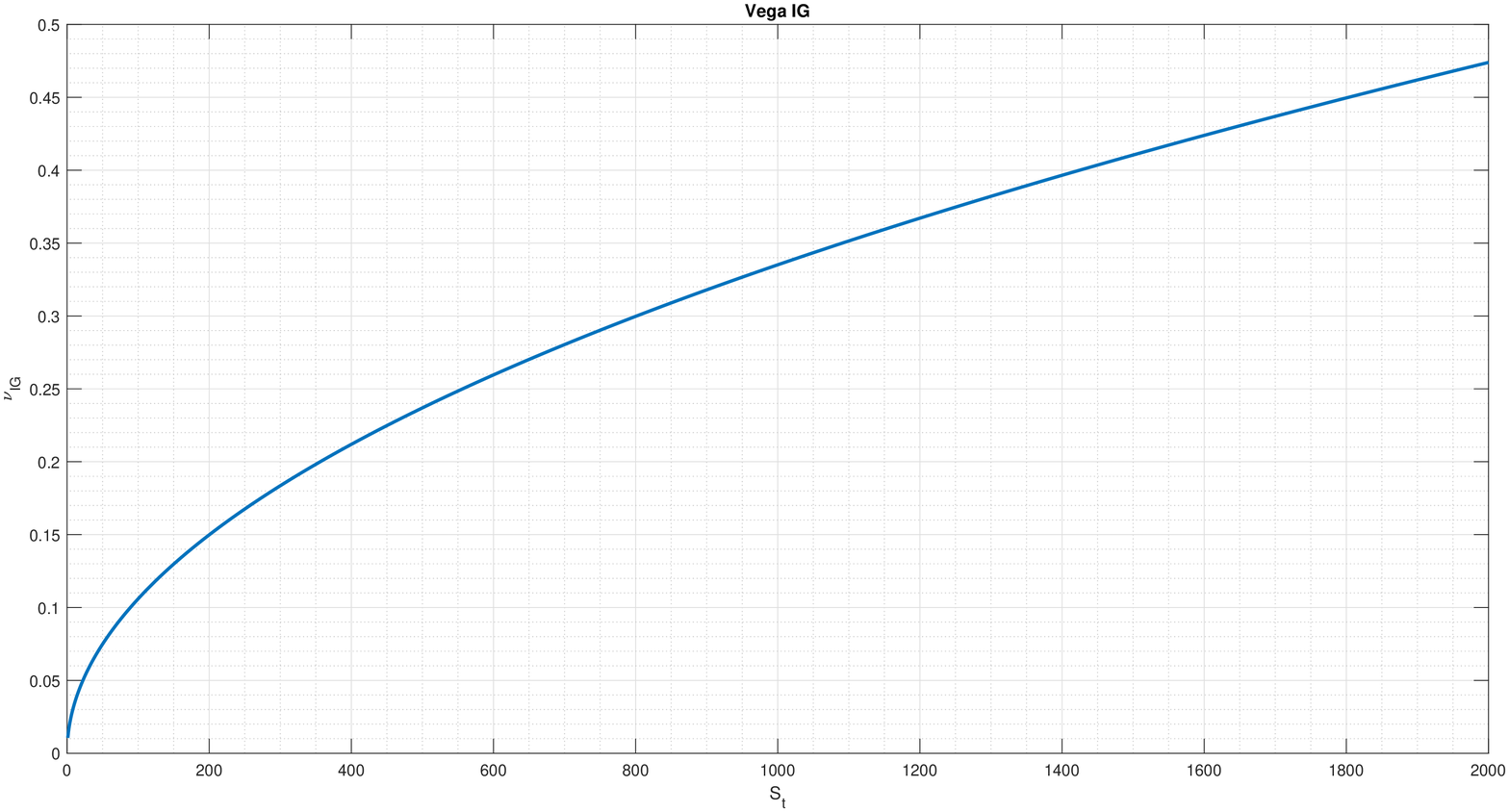}
\end{center}
\endgroup

In the figure we have plotted the Vega corresponding to a 1\% change in volatility $\sigma$, that is $\nu/100$.

\subsubsection{Theta}

$$\Theta_{IG}=V_0\bigg(\frac{r_f}{2}e^{-r_f\tau}-\sqrt{\frac{S_t}{K}}\bigg(\frac{r_f}{2}+\frac{\sigma^2}{8}\bigg)\exp\bigg\{-\bigg(\frac{r_f}{2}+\frac{\sigma^2}{8}\bigg)\tau\bigg\}\bigg)$$

\begingroup
\begin{center}
\includegraphics[scale=.38]{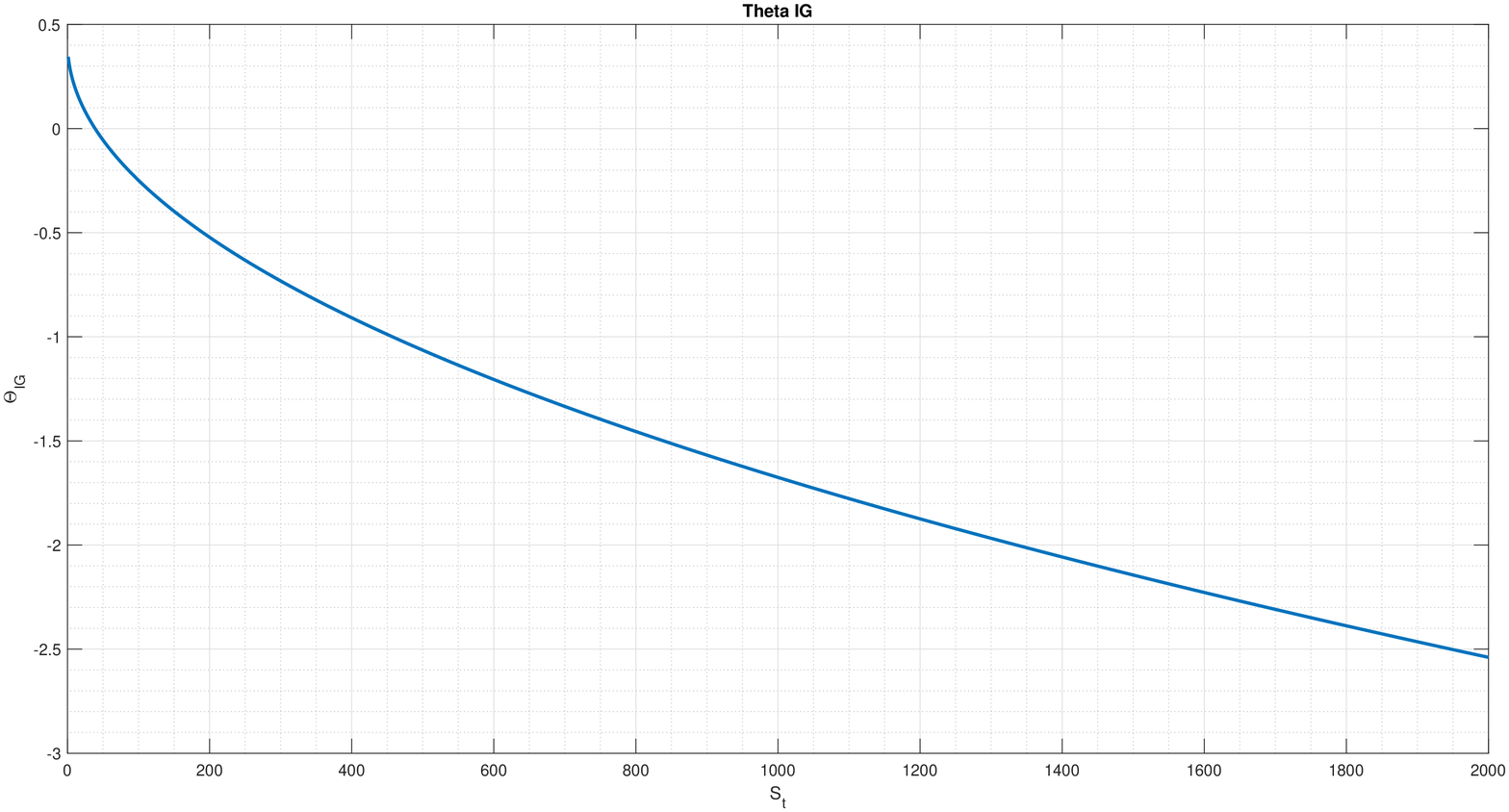}
\end{center}
\endgroup

In the figure we have plotted the daily Theta, that is $\Theta/365$.

\subsubsection{Rho}

$$\rho_{IG}=-\frac{V_0\tau}{2}e^{-r_f\tau}+\frac{V_0\tau}{2}\sqrt{\frac{S_t}{K}}\exp\bigg\{-\bigg(\frac{r_f}{2}+\frac{\sigma^2}{8}\bigg)\tau\bigg\}=\frac{V_0\tau}{2}\bigg(\sqrt{\frac{S_t}{K}}\exp\bigg\{-\bigg(\frac{r_f}{2}+\frac{\sigma^2}{8}\bigg)\tau\bigg\}-e^{-r_f\tau}\bigg)$$

\begingroup
\begin{center}
\includegraphics[scale=.38]{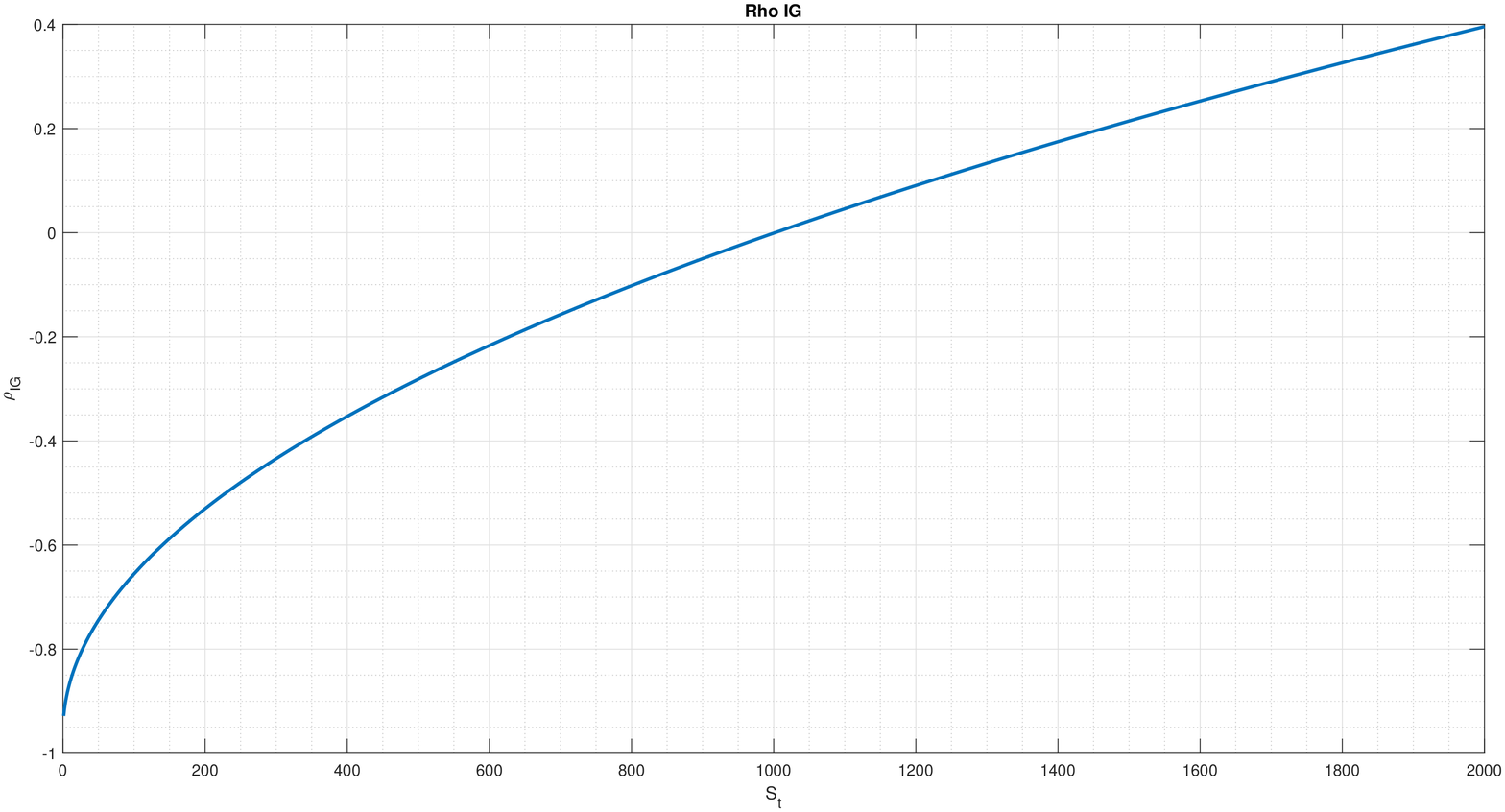}
\end{center}
\endgroup

In the figure we have plotted the Rho corresponding to a 1\% change of $r_x$, that is $\rho/100$.

\subsection{Impermanent Gain as a Hedging Tool}

Imagine an LP provides \$ 10000 of full-range liquidity on Uniswap on the ETH/USDC pair. He then locks his liquidity for 1 year using a liquidity mining platform. ETH price is \$ 1000.
He wants to hedge his position using Impermanent Gain. 

We have:

\begin{itemize}
    \item $V_0=10000$, $S_0 = 1000$;
    \item $r_f=3\%$, $\sigma=70\%$, $\phi = 10\%$;
    \item $T=1$, $\tau = 1$.
\end{itemize}

For the IG, the LP chooses a strike $K$ equal to $S_0$ = 1000 and a maturity of 1 year. We can compute the greeks of this position.

\subsubsection{Delta}

$$\Delta=\Delta_{IG}+\Delta_{LP}=\frac{V_0}{2K}$$

We note that the Delta is a constant determined by the invested capital $V_0$ and the strike $K$. In our case we have $\Delta=5$.

\subsubsection{Gamma}

$$\Gamma=\Gamma_{IG}+\Gamma_{LP}=0$$

Doing this hedge the resulting position is a Gamma-neutral one.

\subsubsection{Vega}

$$\nu=\nu_{IG}+\nu_{LP}=0$$

Doing this hedge the resulting position is a Vega-neutral one.

\subsubsection{Theta}

$$\Theta_{IG}+\Theta_{LP}=V_0r_f\bigg(\frac{1}{2}+\phi T\bigg)e^{-r_f\tau}$$

We note that now Theta doesn't depend on the underlying price $S_t$ and on the volatility $\sigma$.

\subsubsection{Rho}

$$\rho=\rho_{IG}+\rho_{LP}=-V_0\tau\bigg(\frac{1}{2}+\phi T\bigg)e^{-r_f\tau}$$

We note that now Rho doesn't depend on the underlying price $S_t$ and on the volatility $\sigma$.

\vspace{25pt}

As we have seen, for a liquidity provider with locked liquidity, buying Impermanent Gain completely eliminates Gamma and Vega risks as well as significantly reduces Theta and Rho.

\section{Conclusion}

In this paper we analyzed the risks of a liquidity provider with a focus on Impermanent Loss, detailing the position greeks under Black \& Scholes assumptions. We found that a liquidity provider has a positive Delta, negative Gamma and positive Theta, while the Vega is zero; we also demonstrated that locking the liquidity changes the risk profile of a liquidity provider and introduces a negative Vega risk.
Additionally, we introduced Impermanent Gain, a DeFi-native product tailored for liquidity providers' needs and demonstrated how it can be used to eliminate most financial risks related to providing liquidity.

\newpage

\section*{Appendix A}

Recall that:

$$S_T = S_t\exp{\bigg\{\bigg(r_f-\frac{\sigma^2}{2}\bigg)\tau+\sigma W_\tau\bigg\}}\;\;\;\;\;\;\;\;\;\;W_\tau\sim\mathcal{N}(0,\tau)$$

Recall also that given a standard normal distribution $Z\sim\mathcal{N}(0,1)$ we have that:
$$\mathbb{E}\big[\exp\{uZ\}\big]=\exp\bigg\{\frac{u^2}{2}\bigg\}$$

Now we can proceed:

$$\mathbb{E}[\sqrt{S_T}] = \mathbb{E}\bigg[\sqrt{S_t}\exp{\bigg\{\frac{1}{2}\bigg(r_f-\frac{\sigma^2}{2}\bigg)\tau+\frac{\sigma}{2} W_\tau\bigg\}}\bigg] = \sqrt{S_t}\exp\bigg\{\frac{1}{2}\bigg(r_f-\frac{\sigma^2}{2}\bigg)\tau\bigg\}\mathbb{E}\bigg[\exp\bigg\{\frac{\sigma}{2}\sqrt{\tau}Z\bigg\}\bigg]$$

$$\rar\mathbb{E}[\sqrt{S_T}] = \sqrt{S_t}\exp\bigg\{\frac{r_f}{2}\tau-\frac{\sigma^2}{4}\tau\bigg\}\exp\bigg\{\frac{\sigma^2}{8}\tau\bigg\}=\sqrt{S_t}\exp\bigg\{\bigg(\frac{r_f}{2}-\frac{\sigma^2}{8}\bigg)\tau\bigg\}$$

\section*{Appendix B}

Recall that:

$$S_T = S_t\exp{\bigg\{\bigg(r_f-\frac{\sigma^2}{2}\bigg)\tau+\sigma W_\tau\bigg\}}\;\;\;\;\;\;\;\;\;\;W_\tau\sim\mathcal{N}(0,\tau)$$

Recall also that given a standard normal distribution $Z\sim\mathcal{N}(0,1)$ we have that:
$$\mathbb{E}\big[\exp\{uZ\}\big]=\exp\bigg\{\frac{u^2}{2}\bigg\}$$

Now we can proceed:

$$\mathbb{E}[S_T] = \mathbb{E}\bigg[S_t\exp{\bigg\{\bigg(r_f-\frac{\sigma^2}{2}\bigg)\tau+\sigma W_\tau\bigg\}}\bigg] = S_t\exp\bigg\{\bigg(r_f-\frac{\sigma^2}{2}\bigg)\tau\bigg\}\mathbb{E}\big[\exp\big\{\sigma\sqrt{\tau}Z\big\}\big]$$

$$\rar\mathbb{E}[S_T] = S_t\exp\bigg\{r_f \tau - \frac{\sigma^2}{2}\tau\bigg\} \exp\bigg\{\frac{\sigma^2}{2}\tau\bigg\} = S_t e^{r_f\tau}$$

\newpage

\section*{Appendix C}

The following table is a Greeks comparison between the different strategies. Let's first define:

$$\beta \coloneqq \exp\bigg\{-\bigg(\frac{r_f}{2}+\frac{\sigma^2}{8}\bigg)\tau\bigg\}\hspace{50pt}\gamma \coloneqq e^{-r_f\tau}$$

\vspace{10pt}

\begin{center}
\begin{tabular}{ |c|c|c|c| } 
\hline
 \textbf{Greeks} & \textbf{Unlocked LP} & \textbf{Locked LP} & \textbf{Impermanent Gain} \\ 
\hline
 $\Delta$ & $\frac{V_0}{2\sqrt{S_0S_t}}$ & $\frac{V_0}{2\sqrt{S_0S_t}}\beta$ & $V_0\bigg(\frac{1}{2K}-\frac{\beta}{2\sqrt{KS_t}}\bigg)$\\ 
\hline
 $\Delta^{1\%}$ & $\frac{V_0}{2}\sqrt{\frac{S_t}{S_0}}\cdot10^{-2}$ & $\frac{V_0}{2}\sqrt{\frac{S_t}{S_0}}\beta\cdot10^{-2}$ & $V_0\bigg(\frac{S_t}{2K}-\frac{1}{2}\sqrt{\frac{S_t}{K}}\beta\bigg)\cdot10^{-2}$\\ 
\hline
$\Gamma$ & $-\frac{V_0}{4\sqrt{S_0}S_t^{3/2}}$ & $-\frac{V_0}{4\sqrt{K}S_t^{3/2}}\beta$ & $\frac{\partial^2 P_t}{\partial S_t^2}=\frac{V_0}{4\sqrt{K}S_t^{3/2}}\beta$\\
\hline
$\Gamma^{1\%}$ & $-\frac{V_0}{4}\sqrt{\frac{S_t}{S_0}}\cdot10^{-4}$ & $-\frac{V_0}{4}\sqrt{\frac{S_t}{S_0}}\beta\cdot 10^{-4}$ & $\frac{V_0}{4}\sqrt{\frac{S_t}{K}}\beta\cdot 10^{-4}$\\
\hline
$\nu$ & 0 & $-V_0\frac{\sigma\tau}{4}\sqrt{\frac{S_t}{S_0}}\beta$ & $V_0\frac{\sigma \tau}{4}\sqrt{\frac{S_t}{K}}\beta$\\
\hline
$\Theta$ & $\phi\cdot V_0$ & $V_0\bigg(\sqrt{\frac{S_t}{S_0}}\bigg(\frac{r_f}{2}+\frac{\sigma^2}{8}\bigg)\beta+r_f\phi T\gamma\bigg)$ & $V_0\bigg(\frac{r_f}{2}\gamma-\sqrt{\frac{S_t}{K}}\bigg(\frac{r_f}{2}+\frac{\sigma^2}{8}\bigg)\beta\bigg)$\\
\hline
$\rho$ & 0 & $-V_0\bigg(\frac{\tau}{2}\sqrt{\frac{S_t}{S_0}}\beta+\tau\phi T \gamma\bigg)$ & $\frac{V_0\tau}{2}\bigg(\beta\sqrt{\frac{S_t}{K}}-\gamma\bigg)$\\
\hline
\end{tabular}
\end{center}

\newpage

\section*{References}

\begin{enumerate}

    \item Angeris, Chitra, "Improved Price Oracles: Constant Function Market Makers", June 2020, arXiv:2003.10001 [q-fin.TR]

    \item Jensen, Pourpouneh, Nielsen, Ross, "THE HOMOGENEOUS PROPERTIES OF AUTOMATED
    MARKET MAKERS", arXiv:2105.02782 [q-fin.TR]

    \item Park, Andreas, "Conceptual Flaws of Decentralized Automated Market Making" (April 11, 2022). Available at SSRN:
    
    \url{https://ssrn.com/abstract=3805750} or \url{http://dx.doi.org/10.2139/ssrn.3805750}

    \item Clark, Joseph, "The Replicating Portfolio of a Constant Product Market" (March 8, 2020). Available at SSRN:
    
    \url{https://ssrn.com/abstract=3550601} or \url{http://dx.doi.org/10.2139/ssrn.3550601}

    \item Adams, Robinson, Zinsmeister, "Uniswap v2 Core", March 2020, available at:
    
    \url{https://uniswap.org/whitepaper.pdf}

    \item Adams, Keefer, Robinson, Salem, Zinsmeister, "Uniswap v3 Core", March 2021, available at:
    
    \url{https://uniswap.org/whitepaper-v3.pdf}
    
    \item Aigner, Dhaliwal, "UNISWAP: Impermanent Loss and Risk Profile of a Liquidity Provider", June 2021, arXiv:2106.14404 [q-fin.TR]

    \item Elsts, "LIQUIDITY MATH IN UNISWAP V3", 30 September 2021, available at:
    
    \url{https://atiselsts.github.io/pdfs/uniswap-v3-liquidity-math.pdf}

    \item Fukasawa, Masaaki and Maire, Basile and Wunsch, Marcus, Weighted variance swaps hedge against Impermanent Loss (April 27, 2022). Available at SSRN: 
    
    \url{https://ssrn.com/abstract=4095029} or \url{http://dx.doi.org/10.2139/ssrn.4095029}

    \item Black, Scholes, "The Pricing of Options and Corporate Liabilities", \textit{Journal of Political Economy} Vol. 81, No. 3 (May - Jun, 1973), pp. 637-654, The University of Chicago Press

\end{enumerate}

\end{document}